\documentclass[acmsmall,nonacm]{acmart}

\usepackage{cite}
\usepackage{amsmath,amsfonts}
\usepackage{algorithmic}
\usepackage{graphicx}
\usepackage{textcomp}
\usepackage{tabularx}
\newcolumntype{Y}{>{\centering\arraybackslash}X}

\newcolumntype{Z}{>{\hsize=1.3\hsize}X}
\newcolumntype{Q}{>{\hsize=.7\hsize}Y}
\newcolumntype{V}{>{\hsize=.15\hsize}X}
\usepackage{caption}
\usepackage{subcaption}
\usepackage{multirow}
\usepackage{multicol}
\usepackage{makecell}
\usepackage{xcolor}
\usepackage{xurl}
\usepackage{hyperref}

\AtBeginDocument{%
  \providecommand\BibTeX{{%
    \normalfont B\kern-0.5em{\scshape i\kern-0.25em b}\kern-0.8em\TeX}}}

\setcopyright{acmcopyright}
\copyrightyear{2018}
\acmYear{2018}
\acmDOI{XXXXXXX.XXXXXXX}

\acmConference[Conference acronym 'XX]{Make sure to enter the correct
  conference title from your rights confirmation emai}{June 03--05,
  2018}{Woodstock, NY}
\acmPrice{15.00}
\acmISBN{978-1-4503-XXXX-X/18/06}

\begin{document}


\title[Abortion Misinformation on TikTok]{Abortion Misinformation on TikTok: Rampant Content, Lax Moderation, and Vivid User Experiences}


\author{Filipo Sharevski}
\affiliation{%
  \institution{DePaul University}
  \streetaddress{243 S Wabash Ave}
  \city{Chicago, IL}
  \country{United States}}
\email{fsharevs@depaul.edu}

\author{Jennifer Vander Loop}
\affiliation{%
  \institution{DePaul University}
  \streetaddress{243 S Wabash Ave}
  \city{Chicago, IL}
  \country{United States}}
\email{jvande27@depaul.edu}

\author{Peter Jachim}
\affiliation{%
  \institution{DePaul University}
  \streetaddress{243 S Wabash Ave}
  \city{Chicago, IL}
  \country{United States}}
\email{pjachim@depaul.edu}

\author{Amy Devine}
\affiliation{%
  \institution{DePaul University}
  \streetaddress{243 S Wabash Ave}
  \city{Chicago, IL}
  \country{United States}}
\email{adevine@depaul.edu}

\author{Emma Pieroni}
\affiliation{%
  \institution{DePaul University}
  \streetaddress{243 S Wabash Ave}
  \city{Chicago, IL}
  \country{United States}}
\email{epieroni@depaul.edu}







\renewcommand{\shortauthors}{Authors}

\begin{abstract}
The scientific effort devoted to health misinformation mostly focuses on the implications of misleading vaccines and communicable disease claims with respect to public health. However, the proliferation of abortion misinformation following the Supreme Court's decision to overturn \textit{Roe v. Wade} banning legal abortion in the US highlighted a gap in scientific attention to individual health-related misinformation. To address this gap, we conducted a study with 60 TikTok users to uncover their experiences with abortion misinformation and the way they conceptualize, assess, and respond to misleading video content on this platform. Our findings indicate that users mostly encounter short-term videos suggesting herbal ``at-home'' remedies for pregnancy termination. While many of the participants were cautious about scientifically debunked ``abortion alternatives,'' roughly 30\% of the entire sample believed in their safety and efficacy. Even an explicit debunking label attached to a misleading abortion video about the harms of ``at-home'' did not help a third of the participants to dismiss a video about self-administering abortion as misinformation. We discuss the implications of our findings for future participation on TikTok and other polarizing topics debated on social media. 
\end{abstract}




\maketitle

\section{Introduction}
Misinformation, thriving around polarizing topics \citep{ZannettouS}, draws a particular attention in online discourses centered around health issues \citep{Tully}. Alternative health narratives are not a new phenomenon \citep{VoxPop}, but social media's affordances for anonymity, free content creation, and the lack of editorial checks allow for rapid dissemination among users that, in turn, join pro/against camps about ``infant vaccination'' \citep{to-tweet-or-not}, ``COVID-19 mass immunization'' \citep{Weinzierl}, and ``cancer treatments'' \citep{cancer-misinfo} in droves. Health misinformation is not just a harmful pretext for a polarizing social media discourse, but is a real threat for both individual and public well-being \citep{Swire-Thompson-Lazer}. 

In the past, the health misinformation was incited mainly by misleading health research \citep{Dixon}, deceptive interpretations of symptoms \citep{Ghenai}, and contested public health governance decisions \citep{Hart}. In these cases, misinformation appended a fear of either \textit{undesirable} or \textit{unknown} health consequences, prompting people to question long-standing health scientific methods. The response to such misinformation, thus, was complicated by the tendency of people to hold beliefs that align with a persuasive message working on their biases and self-preservation \citep{Ecker-Nature}. In these circumstances, the effort was driven towards prebunking health myths \citep{Lewandowsky}, ``accuracy nudges'' to debunk the misleading health claims \citep{Pennycook-accurate-nudges}, and flagging dangerous health misinformation content on social media \citep{Sharevski-cose}. Some of these interventions did take some of the sting out of the misinformation \citep{Swire-Thompson}, but are far from providing a comprehensive health misinformation containment \citep{Cacciatore}. 

While the scientific community continues to work towards minimizing the adverse effects of ``fear-mongering'' health misinformation \citep{Linden}, a new type of dangerous health misinformation -- appending the lack of \textit{desirable} and \textit{known} health practices -- was abruptly amplified in the immediate aftermath of the US Supreme Court decision to strike down the constitutional right for abortion \citep{Totenberg}. The inability to obtain a legal abortion turned people to search engines and social media to learn how to manage their reproductive decisions and perform safe abortions \citep{Sherman}. Unfortunately, not all information aligned with the National Library of Medicine’s description of abortion and recommendations for safe practices \citep{Ajmal}. 

Many questionable practices including pills, oils, and herbs for inducing abortion flooded social media, both as claims and as an advertisements in users' feeds \citep{Sherman}. Platforms used diverse strategies to mitigate this misinformation: YouTube added ``context labels'' to such abortion content \citep{YouTube}, Twitter decided to promote authoritative abortion information in its Twitter Moments and Events \citep{Kern}, and Meta purportedly blocked questionable abortion treatment advertisements \citep{Meta}. TikTok also stated it removed and labeled videos with abortion misinformation \citep{Keenan}, but many of the questionable home practices aimed to ``cause a miscarriage'' still appeared in users' personal streams \citep{newsguard}. Debunking of abortion misinformation on TikTok followed up \citep{Spencer}, but the slow-in-nature checking and verifying of health-related facts was no match for the rapid spread of videos recommending dangerous abortion remedies. 

Since misinformation in general is ``sticky'', i.e. repeated exposure to false statements make them appear truthful \citep{debunking-handbook}, the lack of systematic response against abortion misinformation at this stage created a situation where ``sticky'' unproven abortion remedies could lead people to attempt unsafe procedures and cause serious bodily harm. While all social media platforms require scrutiny of their abortion misinformation handling, TikTok -- deemed the ``New Google'' for Gen-Z \citep{Gupta} -- draws special attention in this conundrum as pressing reproductive decisions are particularly interesting to the majority of users on this platform. TikTok's status as a platform for social support exchange \citep{Barta} further exacerbates the immediate danger of abortion alternatives as supportive communication adds to ``stickiness'' and internalization of such content among adolescents and young adults \citep{Duggan}.  

Motivated to explore how TikTok users deal with misinformation and alternative abortion narratives, we conducted a study with 60 TikTok users in the United States. First, we obtained a large TikTok dataset to uncover the main themes of abortion misinformation on this platform and get a better sense of how TikTok recommends and moderates such content.
Leveraging the unofficial TikTok-API python \citep{tiktokapi} library, we scraped 8,226 videos with 77,880 hashtags, of which 17,606 were unique in tagging these videos. We  collected the videos using a snowball sampling strategy, starting with scraping three initial hashtags, specifically \#TikTokTaughtMe, \#healthcare, and \#abortion. We selected the first one as it's the \textit{de facto} hashrtag through which a user is ``googling'' short-form TikTok videos for challenges and advice for self-help \citep{newsguard} and the other two as we were focused on health/abortion misinformation in particular.


To report the findings from our study, we review the related work on misinformation and misleading health information on social media in Section \ref{sec:related-work}. Section \ref{sec:abortion-misinfo} provides the broader context of abortion misinformation narratives on social media following the ban on abortion in the United States. Section \ref{sec:study} provides the methodological details of our study and Sections \ref{sec:conceptualzation}, \ref{sec:encounters}, and \ref{sec:dealing} elaborates how participants in our study conceptualize, encounter, and respond to abortion misinformation, respectively. We draw on our findings in Section \ref{sec:discussion} to discuss the implications for both individual and public health as well as social media content moderation of alternative abortion narratives. Finally, Section \ref{sec:conclusion} concludes the paper. 

\section{Health Misinformation Background} \label{sec:related-work}
\subsection{Health Misinformation Narratives}
Health-related rumors and alternative narratives precede the Internet era and were concentrated around health issues with either \textit{unknown} or \textit{undesirable} consequences. In the 1980s, for example, the KGB initiated an information warfare campaign called ``Operation Infektion'' to spread the rumour that HIV/AIDS was a mis-fired American biological weapon in order to undermine the United States' credibility during the Cold War \citep{Boghardt}. In the same time, the tobacco industry in the United States created a ``disinformation playbook'' to systematically distort and downplay the link between the consumption of tobacco and cancer \citep{Reed}. While the intent to mislead was clearly present in these campaigns, the volume and output of the rumors and alternative health narratives was limited to a number of outlets and fabricated publications.  

The Internet and social media changed the landscape by enabling an inordinately high volume and rapid output of health information with varying quality to reach the public \citep{Zielinski}. The health-related rumors and alternative narratives collaterally grew and were amplified to a point where they yielded \textit{uncontrollable} consequences even for \textit{known} and treatable health issues. For example, a poorly designed study in the 1990s that falsely claimed that the measles, mumps, rubella (MMR) vaccine causes autism \citep{surgeon-general} caused such a regression in public immunization that resulted in several measles outbreaks twenty years later on \citep{Hussain}. Rumours about the Ebola and Zika viruses also overshadowed the evidence-based health information and resulted in higher vaccine hesitancy in fear of \textit{undesirable} health consequences and death \citep{Wood-Zika, Pathak}. The vaccine hesitancy, on a global level, achieved a climax during the COVID-19 pandemic with an unprecedented volume and output of COVID-19 related misinformation \citep{Pullan}. 

While the majority of misleading health information on social media focuses on vaccines and communicable diseases, rumors and alternative narratives also spread about cancer, heart disease, and other conditions \citep{Wang}. For example, social media users are more likely to trust and share cancer-related rumours if the rumours are dreadful rather than wishful, and if one has had previous personal experience \citep{Chua}. The \textit{uncontrollable} consequences in these cases are not overall treatment hesitancy but seeking of alternative and unproven treatments about diabetes \citep{Leong}, heart failure \citep{Chen-Hu}, hypertension \citep{Kumar} and psoriasis \citep{Qi-Trang}. Interestingly, in all of these cases of non-communicable health issues, the unsubstantiated claims were promulgated through videos as a particularly influential mode for conveying misleading health evidence (e.g. also used in anorexia and dietary disorders' deceiving messages) \citep{Bessi}.

\subsection{Response to Health Misinformation}
In the context of misleading health claims, misinformation is considered by its opposition to the consensus of what the medical community defines as accurate and evidence-based information \citep{Swire-Thompson-Lazer}. Scholars, in response, have focused the attention of anti-health-misinformation on two main fronts: 1) examining the harms of the misinformation \citep{Loomba, Southwell}; and 2) misinformation prebunking and debunking \citep{Pennycook-Rand-Psych, debunking-handbook, Kirchner};  The harms of health misinformation are reflected in dramatic increase in vaccine hesitancy \citep{Loomba}, pursuing dangerous home therapies (e.g. cancer cleansing, weight loss, virus prevention) \citep{Southwell}, as well as increased hostility toward health workers \citep{Vivek}. 

The goal of ``prebunking'' or forewarning is improving people's ability to spot and resist manipulation techniques commonly used in health misinformation \citep{debunking-handbook}. To this objective, people nowadays are ``innoculated'' against health misinformation by the use of ``accuracy nudges'' \citep{Pennycook-accurate-nudges}, social correction that occurs via peers \citep{Vraga}, or play browser-based games about health myths and facts \citep{Basol}. The ``prebunking'' was shown to be an effective strategy \citep{Lewandowsky}, though in time of social media virality the inoculation effect wanes for users with a conspiracy mentality about \textit{unknown} and \textit{undesirable} health consequences (e.g. the COVID-19 pandemic) \citep{Bertolotti}. 

If this ``innoculation'' is rendered ineffective, ``debunking'' is the next step where verifiable corrections of the falsehoods from credible sources are presented in order to break the illusion of truth \citep{Chan-debunking, Ecker-Nature}. Debunking, as in fact-checking of health misinformation, was shown to give mixed results depending on the perceived credibility and expertise of the sources in science-related contexts  \citep{Zhang-Featherstone}. The perception of credibility and expertise, for example, matters little to people with strong conspiratorial ideation tendencies who tend to mistrust any official source \citep{Lewandowsky-Conspirancy}. As pressing health problems of general public interest are hard not be seen also in a political context, debunking of health misinformation was found to work either when it comes from sources that are perceived to share people's values and worldviews \citep{Brinol}, or when people maintain a science-accepting attitude regardless of their political worldviews \citep{Amazeen}


\subsection{Moderation of Health Misinformation}
In as much as the prebunking and debunking helps in curbing the health misinformation work online, they are nonetheless slow and difficult to scale to the pace, volume, and output of information sharing on social media \citep{Guo-Schlichtkrull}. Platforms, in response, had to turn to automated means of moderating unsubstantiated content and questionable accounts to prevent an ``outbreak'' of misleading information, especially after the meteoric influx of COVID-19 rumors, conspiracies, and falsehoods \citep{Zielinski}. YouTube opted for a soft moderation and decided to apply context labels to video searches for health information that link to credible sources recommended by the National Academy of Medicine \citep{Graham}. Twitter, up to early December 2022, also applied soft moderation in two forms: (i) \textit{interstitial covers}, which obscure the misleading content and require users to click through to see the information; and (ii) \textit{trustworthiness tags}, which appear under the content and do not interrupt the user or compel action \citep{Kaiser, Sharevski-cose}. Meta, the parent company of Facebook and Instagram, did the same in conjunction with hard moderation, taking down prominent accounts that spread COVID-19 misinformation (e.g. Robert F. Kennedy Jr.'s account was blocked after he repeatedly undercut trust in the COVID-19 vaccines) \citep{Chappell}. TikTok followed suit and expanded their soft moderation labeling with trustworthiness tags to content pertaining to eating disorders, health challenges, and alternative medical treatment videos next to misleading COVID-19 videos \citep{TikTok}. 

The response to platform moderation has been, at best, mixed. The interstitial covers provided an adequate ``accuracy nudge'' for users to distance from COVID-19 misinformation posts, but users largely ignored trustworthiness tags \citep{Saltz, context2022}. Further, numerous studies reveal that trustworthiness tags ``backfire'' (i.e. make users believe health misinformation more, not less \citep{Clayton, Nyhan, Ecker, Thorson}). In the context of the COVID-19 pandemic, the tags triggered a ``belief echo,'' manifested as skepticism of adequate mass COVID-19 immunization \citep{Sharevski-cose}. A possible reason for such an unexpected reception of the trustworthiness tags was the asymmetrical nature of soft moderation---the mere exposure to health misinformation often generates a strong and automatic effective response while the tag itself may not generate a response of an equal and opposite magnitude \citep{Gawronski}. This is because the trustworthiness tags often lack meaning, have ambiguous wording, or ask users to find health information themselves (e.g. learn more about COVID-19), which is cognitively demanding and time consuming \citep{Ecker}. 

\subsection{Health Misinformation on TikTok}
TikTok, a social media platform for short-form videos, has rapidly grown in popularity in the last few years \citep{Iqbal}. A central feature of TikTok is the `For You' page, a feed of algorithmically curated videos for the user based on the user's past browsing, viewing, and interactions with videos \citep{Kaye}. Users can also search for videos based on hashtags and, in some cases, sounds. Roughly 75\% of the global users on the platform are age 34 or younger, and every fifth person in the United States visits the platform on a daily basis \citep{Iqbal}. 

TikTok's affordances for viral spread of curated short-form videos and demographic structure \citep{Ling} made the platform particularly interesting for healthcare workers and science communicators creating educational content both based on their speciality and more general advice \citep{Southerton, Zenone}. A review of 331 videos with authoritative COVID-19 information \citep{Li-Guan} showed that anti-stigma/anti-rumor, disease knowledge, encouragement, personal precautions, recognition, and societal crisis management drive platform engagement. Another review of 199 videos with information about chronic obstructive pulmonary disease showed that most of them have a satisfactory scientific background  \citep{Song}. An analysis of obstetrician-gynaecologists (OBGYNs) videos and the associated hashtags and commentary \citep{Southerton} revealed that the health ``educators'' not just convey authoritative sex health information \citep{Stein}, but use it to creatively debunk the related misinformation and misleading treatments. This practice of authoritative health communication as a form of misinformation diffraction also motivated proposals for teaching abortion using TikTok \citep{Duggan}.

Other work on misleading health content on TikTok is scarce and overwhelmingly focuses on COVID-19 health misinformation \citep{McCashin, Zenone}. Basch et al \citep{Basch} analyzed a sample of 72 videos containing the hashtag \#covidvaccine and found that slightly more than half of them discouraged getting the vaccine by showing purportedly adverse vaccine reactions. Baumel et al. \citep{Baumel} analyzed the 100 ``most liked'' videos under each of the \#Pfizer and \#Moderna hashtags and found that 44.2\% and 28.8\% of the comments conveyed misleading negative sentiment against these two COVID-19 vaccines, respectively. Baumel et al. \citep{Baumel-masks} also analyzed TikTok commentary related to masks' effectiveness in combating COVID-19 and found that 45.3\% of commentary using the \#MasksDontWork hashtag contained  misinformation. Analyzing a sample of 1000 videos, Shang et al. \citep{Shang} found that around 22.6\% of the videos contained misleading COVID-19 content, but were shared as much as one with verified COVID-19 information. 

Outside of the COVID-19 theme, O'Sullivan et al. \citep{OSullivan} analyzed 27 TikTok videos containing pediatric urology claims and found that only 22.2\%  contained information that can also be found in official guidelines provided by the European Association of Urology (EAU). Xu et al. \citep{Xu} reviewed 65 TikTok videos with the hashtag \#prostatecancer and found that at least 48\% of them contained explicit prostate cancer misinformation. Zheng et al. \citep{Zheng} study found that the top 100 videos with the \#acne hashtag had seriously misleading information about diagnosis and treatments.

The only study so far analyzing the way misinformation is moderated with warning labels on TikTok focused on COVID-19 content \citep{Ling}. Ling et al. \citep{Ling} collected 41,000 videos that include 26 COVID-19-related hashtags in their description. Through a qualitative analysis, they found out that TikTok likely moderates videos based on hashtags included in the description without an in-depth analysis of the content. Ling et al. learned that this moderation strategy led to a large false positive rate -- about a quarter of the videos with a misinformation warning label did not contain content related to COVID-19. The study also found a 7.7\% false negative rate where videos with actual COVID-19 misinformation did not include warning labels.

\section{Abortion Misinformation Context} \label{sec:abortion-misinfo}
Prior studies have shown that 70.1\% of women obtain information regarding abortion from the Internet \citep{Littman}. Abortion misinformation online, thus, takes many forms and users generally have difficulties discerning inaccuracies in the related alternative narratives \citep{Patev}. Bessett et al. \citep{Bessett} presented 586 participants with five vignettes of abortion misinformation -- safety, breast cancer, infertility,
mental health risk, and legality of abortion -- and found that only 4\% of participants were able to correctly identify all of  vignettes as misinformation while 73\% pointed to two or fewer vignettes as inaccurate.  

Common abortion misinformation topics that have been studied are the increased risk of breast cancer, future infertility, depression/anxiety, and post-traumatic stress \citep{Patev}. This misinformation is spread through multiple sources, including state-mandated ``\textit{Women's Right to Know}'' documentation that providers must supply before a woman can consent to having an abortion \citep{Berglas}, despite official guidance from the National Academies of Sciences, Engineering, and Medicine \citep{NAS}. The Guttmacher Institute found that two states inaccurately include a link between abortion and an increased risk of breast cancer, 19 states link abortion to future infertility, and eight states link abortion to negative emotional and psychological responses \citep{Guttmacher}. Kern and Reader \citep{Kern} reported that abortion misinformation specifically related to an ``abortion reversal pill'' increased on Facebook from 20 interactions on June 23 to 3,500 interactions on June 24 2022, the day after the Supreme Court decision to overturn \textit{Roe v. Wade}. Godoy \citep{Godoy} also reported that following the Supreme Court ruling, Spanish-language abortion misinformation was deliberately designed to galvanize voters in Latino communities across the US. 

Abortifacient herbs -- purportedly providing the ability to induce a spontaneous miscarriage -- form the majority of post-\textit{Roe v Wade} misinformation \citep{Swenson}. The toxicity of abortifacient herbs has been widely studied as shown in Table \ref{tab:herbs} but there is little literature and few studies related to the topic of ``herbal abortions'' \citep{Johnson-Arbor}. Most existing studies were done in countries where abortion was not legal until recently. Abortion did not become legal in Uruguay until 2012 \citep{Makleff}, for example, and a 2003 study found that the Montevideo Poison Centre had 86 cases of ingestion of herbal infusions with abortive intent from 1986 to 1999 \citep{Ciganda}. In the United States, misinformation surrounding ``herbal abortions'' in viral videos on TikTok has increased dramatically after legal abortion was overturned \citep{Case}. The consequences of these viral misinformation videos already brought several people to the emergency rooms seeking critical lifesaving treatment, making active prebunking and debunking by qualified health professionals imperative \citep{RubinR}. 

\begin{table}[htbp]
\renewcommand{\arraystretch}{1.3}
\footnotesize
\caption{Herbal Abortifacients and Side Effects}
\label{tab:herbs}
\centering
\begin{tabularx}{0.8\linewidth}{|lX|}
\Xhline{3\arrayrulewidth}
\textbf{Common Name} & Side Effects \\\Xhline{3\arrayrulewidth}
\textbf{Black Cohosh} & Hepatoxicity \citep{Enbom} \\\Xhline{3\arrayrulewidth}
\textbf{Blue Cohosh} & Nicotinic toxicity: tachycardia, hypertension, headaches, abdominal pain, vomiting, muscle weakness and fasciculations, seizures, and coma \citep{Rao} \\\Xhline{3\arrayrulewidth}
\textbf{Eastern Daisy Fleabane} & Insufficient evidence on the safety and effectiveness as an abortifacient agent \citep{Thoms} \\\Xhline{3\arrayrulewidth}
\textbf{Mugwort} & Vomiting, hypertension, confusion, respiratory distress, coma, and seizures \citep{Missouri} \\\Xhline{3\arrayrulewidth}
\textbf{Parsley} & Abdominal pain, vomiting, genital hemorrhage, anemia, jaudice \citep{Ciganda}, internal bleeding, convulsions, and death \citep{Cassella} \\\Xhline{3\arrayrulewidth}
\textbf{Pennyroyal} & Gastrointestinal upset, fainting, intestinal bleeding, seizures, hepatomegaly or injury, multiple organ failure, coma, cardiac arrest, and death \citep{Berglas-Pennyroyal, Gupta} \\\Xhline{3\arrayrulewidth}
\textbf{Rue} & Vomiting, liver damage, anemia, tremors, respiratory distress, multiple organ failure, and death \citep{JohnsonA} \\\Xhline{3\arrayrulewidth}
\end{tabularx}
\end{table}

\section{Misinformation and Alternative Abortion Narratives on TikTok} \label{sec:study}
\subsection{Research Questions}
The volume and output of abortion misinformation naturally prompted health experts, by themselves, to dispel the inaccuracies related to herbal abortions, abortion pills, and abortion side-effects directly on social media \citep{Serrano}. This effort is by all means needed, but is likely not to be sufficient to prevent an ``outbreak'' of unsafe decisions about people's reproductive health in the long run. An intuitive response, then, would be a systematic prebunking and debunking of abortion misinformation in coordination with moderation of related content on social media. Based on past experiences with health misinformation, however, such a response leaves little room for nuanced explorations of how people \textit{engage on their own} with abortion misinformation in the first place \citep{Southerton}. 

As TikTok has been identified as a ``hotbed of abortion misinformation'' \citep{Grierson}, such an exploration would be beneficial to the ongoing response of removing and moderating misleading content on TikTok \citep{Keenan} as it will provide knowledge on how users conceptualize, encounter, assess, and respond to abortion falsehoods. So far, such knowledge is scarce and only provides glimpses on how users conceptualize misinformation encountered on the traditional social media platforms \citep{folk-models}. To address this knowledge gap, we set to conduct a study that aimed to answer the following research questions:   
\vspace{0.5em}
\begin{enumerate}
\itemsep 0.5em
    \item \textbf{RQ1:}\ \textit{Concept}: How do social media conceptualize misinformation on TikTok (definition, origins, targets, and purpose)?
    \item \textbf{RQ2:}\ \textit{Encounters}: What encounters with abortion misinformation users had so far on TikTok and how they dealt with it? 
    \item \textbf{RQ3:}\ \textit{Response}: What strategies users employ in assessing and responding to various abortion misinformation content on TikTok?
\end{enumerate}

\subsection{Dataset}
Preliminary, we set to collect a dataset of abortion misinformation on TikTok in the immediate period after the overturn  of Roe vs Wade, up till the end of November 2022, as shown in Table \ref{tab:dataset}. We leveraged the unofficial TikTok-API python library \citep{tiktokapi} to scrape 8,226 videos, which we collected using a snowball sampling strategy, starting with scraping three initial hashtags, specifically \#TikTokTaughtMe, \#Healthcare, and \#Abortion. Due to the limitations of the API, each search returned no more than 300 videos. To continue collecting hashtags, we searched for each of the hashtags that were associated with each of the three seeding hashtags above, effectively performing a snowballing sampling of the TikTok's base with abortion-related short-form videos. 

\begin{table}[htbp]
\renewcommand{\arraystretch}{1.3}
\footnotesize
\caption{TikTok Abortion Hashtag Dataset}
\label{tab:dataset}
\centering
\begin{tabularx}{0.8\linewidth}{|YY|}
\Xhline{3\arrayrulewidth}
\textbf{Attribute} & \textbf{Value} \\\Xhline{3\arrayrulewidth}
Total Number of Posts & 8,226 \\\Xhline{3\arrayrulewidth}
Number of Hashtags & 77,880 \\\Xhline{3\arrayrulewidth}
Unique Hashtags & 17,606 \\\Xhline{3\arrayrulewidth}
\end{tabularx}
\end{table}

From here, we vectorized the hashtags using Scikit-Learn's CountVectorizer \citep{sklearn} to create a dense boolean array of 1,754 tokens -- character unigrams, bigrams, and trigrams -- that appeared in 0.01 - 99\%  of hashtag samples. We then identified the closest hashtags to a given input hashtags using Minkowski distance, or $||x||_p$ where $x$ is the difference between an searched vector and the saved vectors from our hashtag dataset, and $p$ is a scalar that we selected. To identify $p$, we reviewed kernel density estimate plots of the distances to several searched hashtags and identified the most expected bimodal distribution, with a smaller left distribution of relevant hashtags, and a larger right distribution of less relevant hashtags. We settled on an ideal $p$ of $2$, which is $||x||_2$, or euclidean distance.

Using these hashtag representations, we were easily able to identify perturbations in hashtags that might otherwise be moderated by TikTok \citep{Keenan}. For example, searching the representations for hashtags like \#selfharm highlighted the existence of \#s\^{e}lfh\^{a}rm, and \#abortion revealed \#abotion and \#anortion, as well as longer hashtags like \#abortionishealthcare and \#abortionishealth\c{c}are. A search with regular terms and hashtags like ``\#abortifacient'' indeed does not present any videos tagged as such, but following our dataset analysis above, we discovered that a small change in the spelling -- \#ab0rtifacient, for example -- unveils a lot of abortion videos that promote abortifacient solutions for miscarriage. 

\begin{figure*}[htp]

\centering
\includegraphics[width=.25\textwidth]{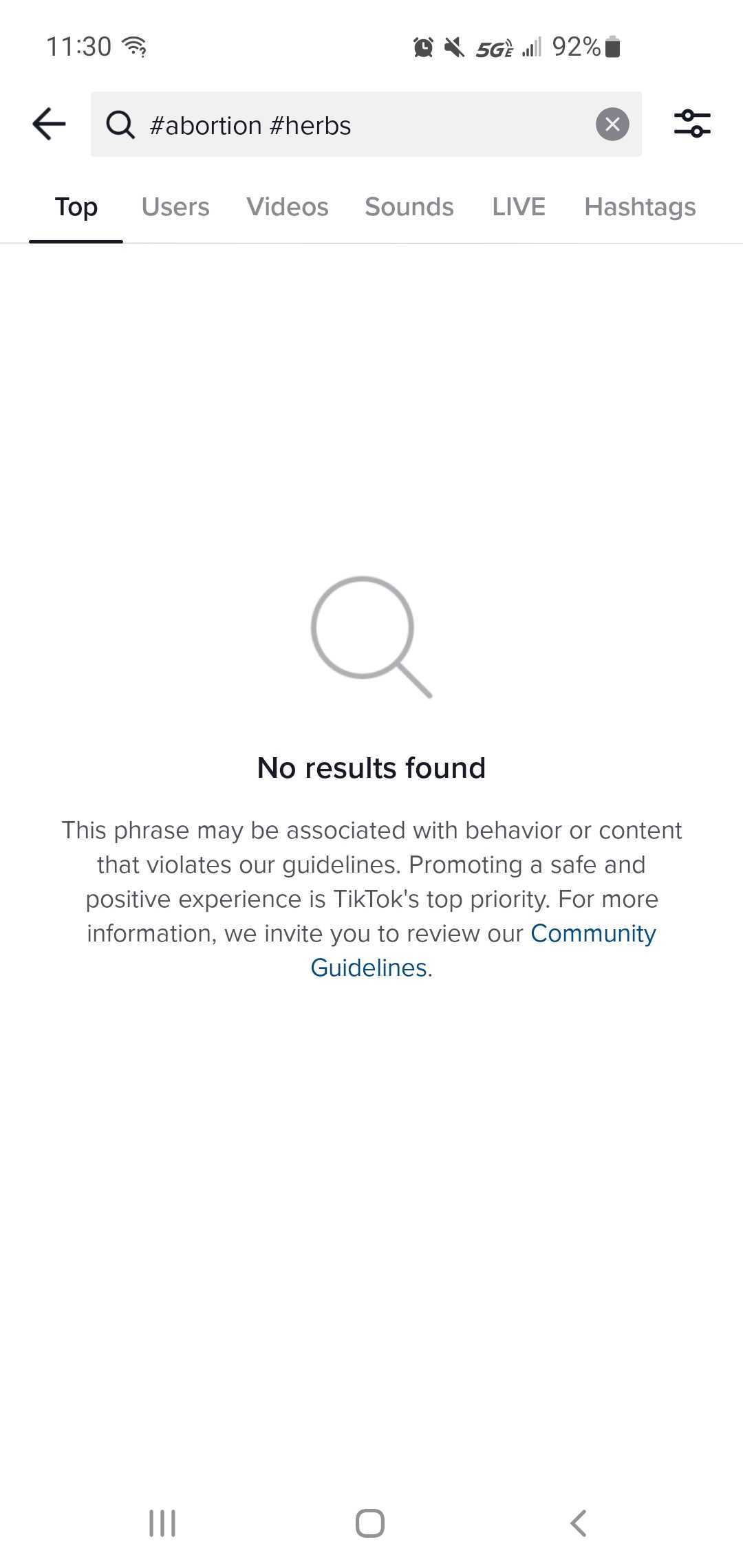}\hfill
\includegraphics[width=.25\textwidth]{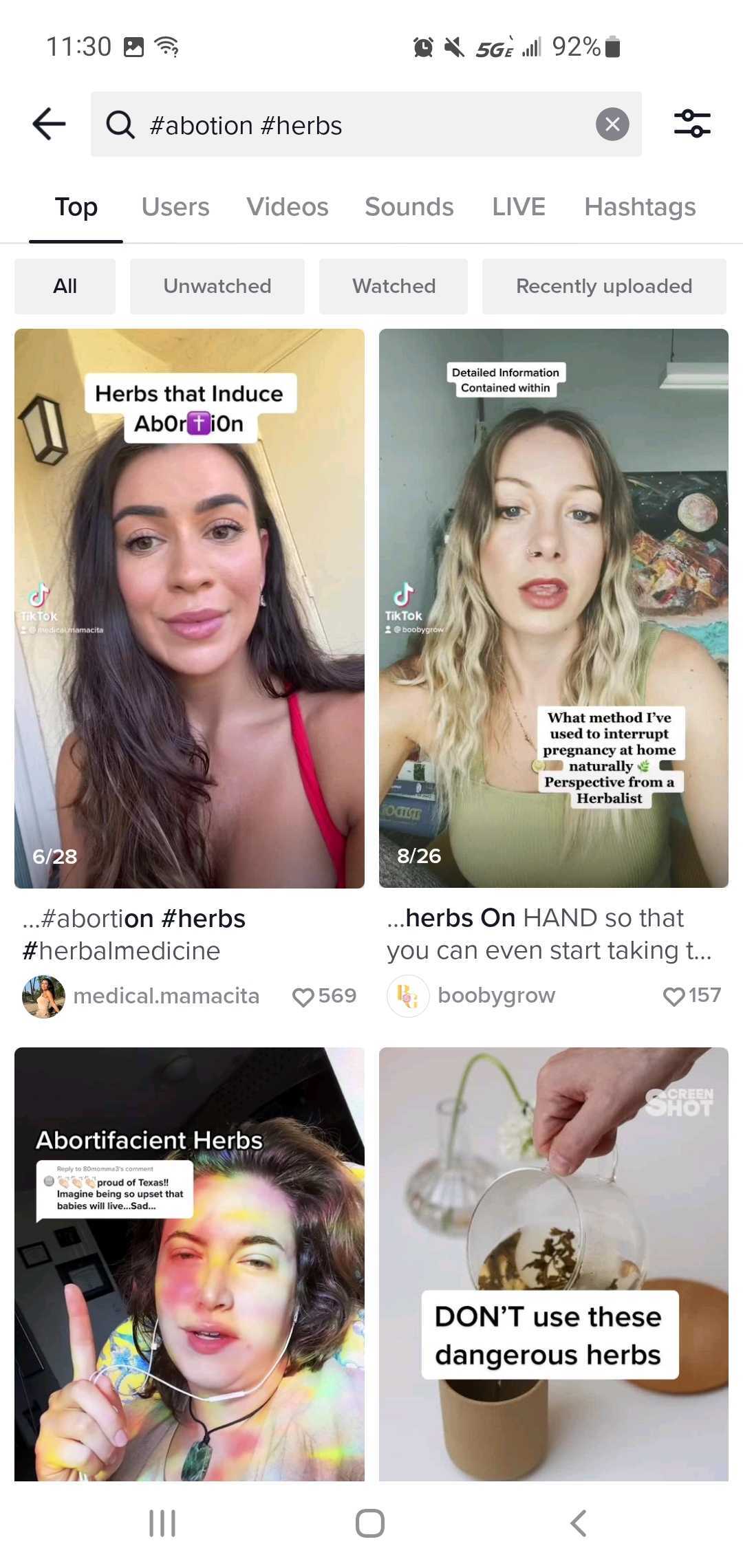}\hfill
\includegraphics[width=.25\textwidth]{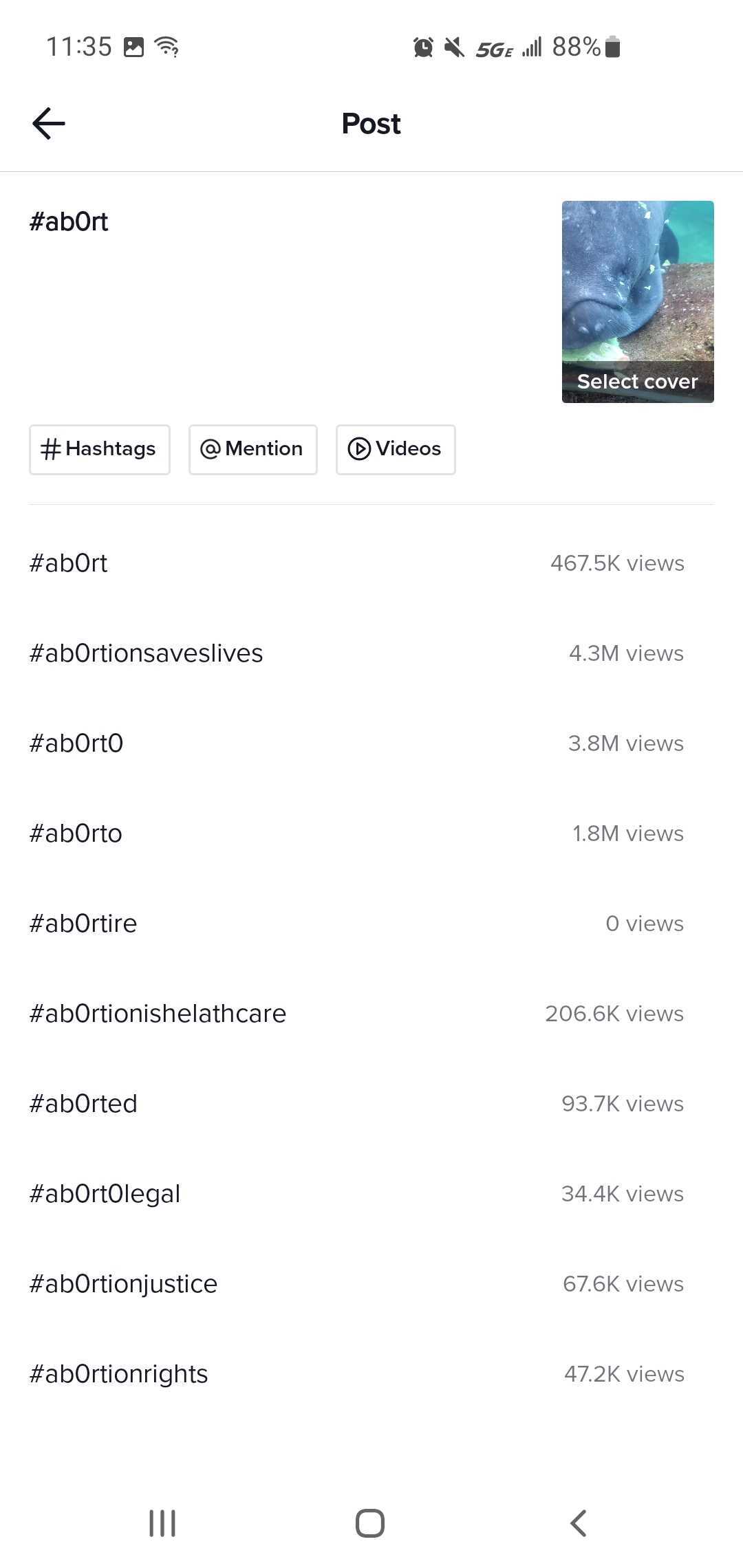}

\caption{The images above demonstrate how, while a search for ``\#abortion \#herbs'' does not return any videos due to the TikTok's guidelines for harmful content \citep{Keenan}, a search for ``\#abotion \#herbs'' not only returns videos with similar typos, it also returns videos with the original ``\#abortion \#herbs'' spelled correctly. Additionally, when posting a video, TikTok suggests additional hashtags, any of which can be searched to find additional hashtags, like \#ab0rtionishelathcare.}
\label{fig:hashtagsearch}

\end{figure*}

Many of these videos were not necessarily were tagged with the exact search hashtag and may even be tagged with the original  ``\#abortifacient.'' One could argue that an ordinary users might not know what hashtags exist, but from the video posting functionality, a list of suggested hashtag completions provide additional variations, and the number of videos with each variation. As such, even an incomplete, suggestive hashtag search on TikTok brings seemingly obscured tags for abortion misinformation, as exemplified in Figure \ref{fig:hashtagsearch} Using these built in features, we quickly identified dozens of videos that described methods for ``at-home'' abortions as candidates for misleading claims we wanted to test in our study.




\subsection{Sample}
The analysis of the information in our dataset, given in section \ref{sec:dealing}, helped us identify the main themes of abortion misinformation content on TikTok in the aftermath of \textit{Roe vs Wade} decision. As we were interested in better understanding how actual users deal with this content, we obtained approval from our Institutional Review Board (IRB) to conduct an exploratory survey (the questionnaire is provided in the \hyperref[sec:survey]{Appendix}) with a sample of TikTok users ages 18 and above in the United States. We used Prolific for recruitment and after we consolidated the responses we obtained through Qualtrics, we ended with a sample of total of 60 participants. The responses were anonymous, and the survey allowed users to skip any question they were uncomfortable answering, taking around 25 minutes to complete it. Participants were offered a compensation rate of \$5 each. The demographic structure of our sample is given in Table \ref{tab:demographics}.

\begin{table}[htbp]
\renewcommand{\arraystretch}{1.3}
\footnotesize
\caption{Sample Demographic Distribution}
\label{tab:demographics}
\centering
\aboverulesep=0ex 
   \belowrulesep=0ex 
\begin{tabularx}{0.8\linewidth}{|Y|}
\Xhline{3\arrayrulewidth}
\toprule
 \textbf{Gender} \\\Xhline{3\arrayrulewidth}
\midrule
\footnotesize
\vspace{0.2em}
    \hfill \makecell{\textbf{Female} \\ 44 (73.33\%)} 
    \hfill \makecell{\textbf{Male} \\ 15 (25\%)} 
    \hfill \makecell{\textbf{Non-cisgender} \\ 1 (1.67\%)} \hfill\null
\vspace{0.2em}
\\\Xhline{3\arrayrulewidth}
\midrule
 \textbf{Age} \\\Xhline{3\arrayrulewidth}
\midrule
\footnotesize
\vspace{0.2em}
\hfill \makecell{\textbf{[18-20]}\\ 5 (8.33\%)} 
    \hfill \makecell{\textbf{[21-30]} \\ 33 (55\%)} 
    \hfill \makecell{\textbf{[31-40]}\\ 12 (20\%)} 
    \hfill \makecell{\textbf{[41-50]} \\ 6 (10\%)}
    \hfill \makecell{\textbf{[51-60]} \\ 4 (6.67\%)}
    \hfill \makecell{\textbf{[61+]} \\ 0 (0\%)}
\vspace{0.2em}
\\\Xhline{3\arrayrulewidth}
\midrule
 \textbf{Political leanings} \\\Xhline{3\arrayrulewidth}
\midrule
\footnotesize
\vspace{0.2em}
    \hfill \makecell{\textbf{Left} \\ 38 (63.33\%)} 
    \hfill \makecell{\textbf{Moderate} \\ 14 (23.33\%)} 
    \hfill \makecell{\textbf{Right} \\ 5 (8.33\%)} 
    \hfill \makecell{\textbf{Apolitical} \\ 3 (5\%)} \hfill\null 
\vspace{0.2em}
\\\Xhline{3\arrayrulewidth}
\midrule
 \textbf{Highest Level of Education Completed} \\\Xhline{3\arrayrulewidth}
\midrule
\footnotesize
\vspace{0.2em}
    \hfill \makecell{\textbf{High school} \\ 12 (20\%)} 
    \hfill \makecell{\textbf{College} \\ 43 (71.67\%)} 
    \hfill \makecell{\textbf{Graduate} \\ 5 (8.33\%)} \hfill\null 
\vspace{0.2em}
\\\Xhline{3\arrayrulewidth}
\bottomrule
\end{tabularx}
\end{table}


\subsection{Method and Analysis}
Participants were provided an open ended qualitative survey through Prolific that provided a list of questions and a predetermined set of TikTok videos we selected from our dataset. We singled out seven videos in total from our dataset that contained  abortion misinformation already debunked by the time of our study \citep{Spencer}. We used the input on general abortion misinformation from Table \ref{tab:herbs}, information from authoritative verifiable sources \citep{NAS}, and verbatim misinformation terms from two fact-checking articles \citep{Cercone, Thoms} as a selection criteria for videos promoting the use of herbal abortifacients. We also chose to focus only on ``at-home abortion remedies'' as explicit health misinformation \citep{Southerton} and not alternative abortion narratives involving ``religion'' or ``political contextualizaiton'' to avoid bias and expressive responding \citep{Berinsky}.  

We wanted to have as many varying modalities, formats, and creators in our selection as possible, therefore we selected two videos that contained only text and five videos featuring the creator of the content. Six of the selected videos were created by women and one was created by an individual who identifies as transgender in their profiles. The creators of the videos were ethnically diverse, consisting of individuals who identify in their profile or other videos as White, Black, North American Indigenous, and Native Hawaiian or Pacific Islander. We must note that TikTok, in response to the increased scrutiny about their lax handling of health misinformation \citep{Keenan}, claims to regularly remove misleading abortion content so there is a possibility that our dataset was considerably restricted for our particular selection.

Participants were asked to describe their experience with encountering misinformation on TikTok. Next, we asked participants to provide their opinions on where misinformation comes from, what purpose misinformation serves on social media, and who creates and benefits from it. Participants were then asked to further elaborate how they determine a certain social media post is misinformation, and what tactics they employ when dealing with misinformation.


In reporting the results, we utilized as much as possible verbatim quotation of participants' answers, emphasized in ``\textit{italics}'' and with a reference to the participant as either \textbf{PXYZ\#} or [\textbf{PXYZ\#}], where \textbf{P} denotes \textbf{participant}, \textbf{X} denotes the \textbf{number} of the participant in the sample (ordered by the time of participation), \textbf{Y} denotes their \textbf{gender} identity (\textbf{F} - female, \textbf{M} - male, \textbf{NC} - non-cisgender), \textbf{Z} denotes their \textbf{political} identity (\textbf{L} - left-leaning, \textbf{M} - moderate, \textbf{R} - right-leaning; \textbf{A} - apolitical), and \textbf{\#} denotes the upper bound of their \textbf{age bracket}. For example, \textbf{P16FL30} refers to \textbf{participant 16}, \textbf{female}, \textbf{left-leaning}, \textbf{age bracket} \textbf{[21-30]}.


\section{Misinformation Conceptualization on TikTok} \label{sec:conceptualzation}


\subsection{Definition} 
First, we asked our participants to define misinformation in their own words. Exactly half the sample provided a definition that did not include any intention in the production or dissemination of questionable content, along the lines of the \textit{misinformation} definitions outlined in \citep{Wu}. All of these participants conceptualized falsehoods through the \textit{inherently fallacious information} mental model of misinformation on social media described in \citep{folk-models}. For example, \textbf{P31FL30} defined it as  ``\textit{untrue/unsubstantiated statements being presented as fact},'' \textbf{P13FR30} as ``\textit{incorrect, skewed, or communicated incorrectly},'' and \textbf{P27MM20} as simply ``\textit{false information}.'' In this half of the sample, 18 (60\%) of the participants identified as left-leaning, 3 (10\%) as right-leaning, 8 (26.67\%) as moderate, and one (3.33\%) as apolitical. 


The other half of our sample expressed intentionality as an additional quality of misinformation, \textit{de facto} referring to \textit{disinformation} instead \citep{ZannettouS}. Using the folk models of misinformation on social media \citep{folk-models}, more than half, 20 (66.67\%), of the participants conceptualized misinformation as \textit{out-of-context narratives}, for example, \textbf{P36FL40} stated that misinformation is ``\textit{is intentionally, either by using wrong information or leaving out context, misleading to the people reading it}.'' The next most popular folk model 6 (20\%) was \textit{external propaganda} and the participants pointed to ``\textit{intentional spread of misleading information to stir an emotion or to further promote a system, product, or person}'' [\textbf{P5FL30}]. The remaining 4 (13.33\%) participants conceptualized misinformation as \textit{political (counter)argumentation} pointing to cases where ``\textit{a journalist or news source provides false information to persuade you in one political direction}'' [\textbf{P47MR30}]. Here, the older participants were, the more they saw an intention in the spread of misinformation. For example, \textbf{P50FL60} placed the intentionality where ``\textit{videos get edited and changed, and convincing memes with cherry picked facts are created as part of misinformation that has been used extensively in politics and the pandemic}.''

\subsection{Origins}
Three quarters or 45 of the participants in our sample felt that misinformation on TikTok came directly from a \textit{creator of the TikTok video}. In the view of \textbf{P20FL40}, misinformation on TikTok is brought by ``\textit{people who are trying to gain clout, or get numerous views}.'' \textbf{P19FL30} went further and reckoned that ``\textit{misinformation can come from the creators' own consumption of misinformation, or a creators' misinterpretation of information, or a creators' attempt to sell something or an idea to influence others/gain attention}.'' The creators of Tiktok content, in the view of \textbf{P27MM20}, ``\textit{are people who doesn't care about misinformation but more about views and attention}''.

The remaining 25\% pointed to the \textit{``other'' side} of a polarized debate or issue i.e. ``\textit{people on both the left and right who want to increase views, as well as institutions and political groups with agendas to create misinformation}'' [\textbf{P50FL60}]. \textbf{P2FL50}, seeing misinformation as out-of-context narrative, felt that it ``\textit{comes from a variety of places such as Republicans, Russia or China}'' and \textbf{P12FL60} seconded the impression of external interference ``\textit{directly from a bad-actor or a big-mouth source such as Fox News}. \textbf{P50FL60}, using the political (counter)argumentation model of misinformation, directly accused the GOP for ``\textit{catering misinformation to low information and low IQ people who will believe anything they get told because GOP knows they can't fool the science/college crowd}.''

\subsection{Targets}
Half of our sample felt that the targets of misinformation are ``\textit{vulnerable people who do not know how to research and form their own opinions}'' [\textbf{P7FM30}], specifically ``\textit{Younger people, older people, or more easily-influenced crowds, which are the people who are not likely to fact check a claim}'' [\textbf{P40FL30}]. \textbf{P50FL60} expanded this list to include people ``\textit{in areas that have low instances of college education and high poverty areas; who are lower income and very religious; who are already suffering themselves and see anyone who gets ahead as a threat to them; and who have very little access to help so they resent people}.'' The other half felt that ``\textit{anyone and everyone can be a target of misinformation}'' [\textbf{P44FL30}]. \textbf{P5FL30} described the targets of misinformation on TikTok to be from ``\textit{All ages, races, sexualities, and backgrounds are targets of misinformation because the algorithm brings it to your `For You page'}.''

\subsection{Purpose}
20 (33.33\%) of our participants explicitly indicated that the purpose of misinformation on TikTok is for profit. The profit was assigned either to ``\textit{politicians and large corporations who either make money off what evolves from misinformation campaigns or who benefit politically and financially from legislation enacted when bad actors are elected to government offices}'' [\textbf{P12FL60}] or to content creators themselves as ``\textit{they get paid from the views, and there's probably some devout followers to these people which give them a recurring income from just watching the videos every time they post}'' [\textbf{P15ML30}]. Implicit gains, such as ``\textit{engagement boosts}'' [\textbf{P1MR50}] that ultimately lead to profit per the TikTok participation model \citep{Kaye}, was the purpose that 14 (23.33\%) of our participants identified behind the spread of misinformation on TikTok. They identified ``\textit{creators and influencers looking to gain followers and views}'' [\textbf{P1MR50}] and ``\textit{people who doesn't [sic] care about misinformation but more about views and attention}'' [\textbf{P27MM20}].

Misinformation as a political ammunition was the purpose identified by 13 (21.67\%) of our participants. \textbf{P31FL30} indicated that the purpose of misinformation is ``\textit{political influence feeding into distrust of science and government}'' and \textbf{P59FL30} felt the misinformation on TikTok is brought ``\textit{to divide people further and continue to build up the conservative party}.'' 5 (8.33\%) of participants felt that misinformation was to ``stir the pot'' on TikTok, i.e. ``\textit{foreign agency targeting the US or groups within the US that want superiority}'' [\textbf{P41FM40}]. Videos created by ``\textit{trolls make up a portion of deliberate misinformation}'' [\textbf{P38FL30}] to  ``\textit{gets a rise out of someone},'' in the view of \textbf{P5FL30}. A subgroup of 8 (13.33\%) participants, indicated that ``\textit{no one}'' [\textbf{P10MA40}] benefits from misinformation on TikTok, both in a short-run and ``\textit{ultimately, in the long-run}'' [\textbf{P58FM30}].

\section{Abortion Misinformation Encounters on TikTok} \label{sec:encounters}
\subsection{Encounters}
Exactly half of the sample indicated they have seen abortion misinformation on TikTok prior to the study. The misleading content mostly consisted of ``\textit{videos like these claiming at home abortion remedies}'' [\textbf{P25FL20}] but also included ``\textit{misinformation rooted in religion -- churches show videos of full term pregnancies being ripped apart by limbs from wombs}'' [\textbf{P50FL60}]. Participants also indicate they were seeing politically contextualized abortion narratives ``\textit{on both sides of the political spectrum}'' [\textbf{P7FM30}] to either ban or allow ``\textit{birth control as well as contraceptives provided by the government}'' [\textbf{P24FR20}]. Participants also indicated that they see ``\textit{people who are Pro Life on TikTok that spread all kinds of rumors and lies about abortion all the time}'' [\textbf{P5FL30}] ``\textit{mostly to cause fear}'' [\textbf{P49Fl30}].

\subsection{Response}
About half of the participants indicated that abortion misinformation invoked negative emotions in them. Participants stated that the ``\textit{videos in all just made me sad}'' [\textbf{P6FA30}], that they were ``\textit{very disturbed by this abortion misinformation}'' [\textbf{P13FR30}], and ``\textit{disappointed that people are making content like this}'' [\textbf{P15ML30}]. Some of them said that their ``\textit{first response was shock that someone would even think this}'' [\textbf{P24FR20}], that and it is ``\textit{worrying that this type of misinformation is being shared because it can be dangerous}'' [\textbf{P39-FL30}]. Participants in our sample felt ``\textit{anger, resentment}'' [\textbf{P42FL30}] and ``\textit{disgust that people will believe anything they see and try it}'' [\textbf{P56FM50}]. The other half indicated they were mostly ``\textit{intrigued and wanted to know if anything in these herbal videos was true or not}'' [\textbf{P41FM40}]. 

In response to abortion misinformation content on TikTok, our participants said they ``\textit{did not engage because it only further spreads the misinformation}'' [\textbf{P8FL20}] or ``\textit{just ignored it}'' [\textbf{P52MM40}]. Some participants indicated they took action on the video by doing ``\textit{research on abortion and take what I gather on TikTok with a grain of salt unless it is information spread by an actual health professional}'' [\textbf{P26FL30}]. There were also participants that ``\textit{blocked the creators that spread the misinformation}'' [\textbf{P20FL40}], ``\textit{reported these videos for spreading false information}'' [\textbf{P49Fl30}], or ``\textit{Liked comments pointing out the abortion falsehoods}'' [\textbf{P31FL30}]. 

\section{Response to Abortion Misinformation on TikTok} \label{sec:dealing}

\subsection{Post \#1}
\begin{figure}[htbp]
  \centering
  \includegraphics[width=0.35\linewidth]{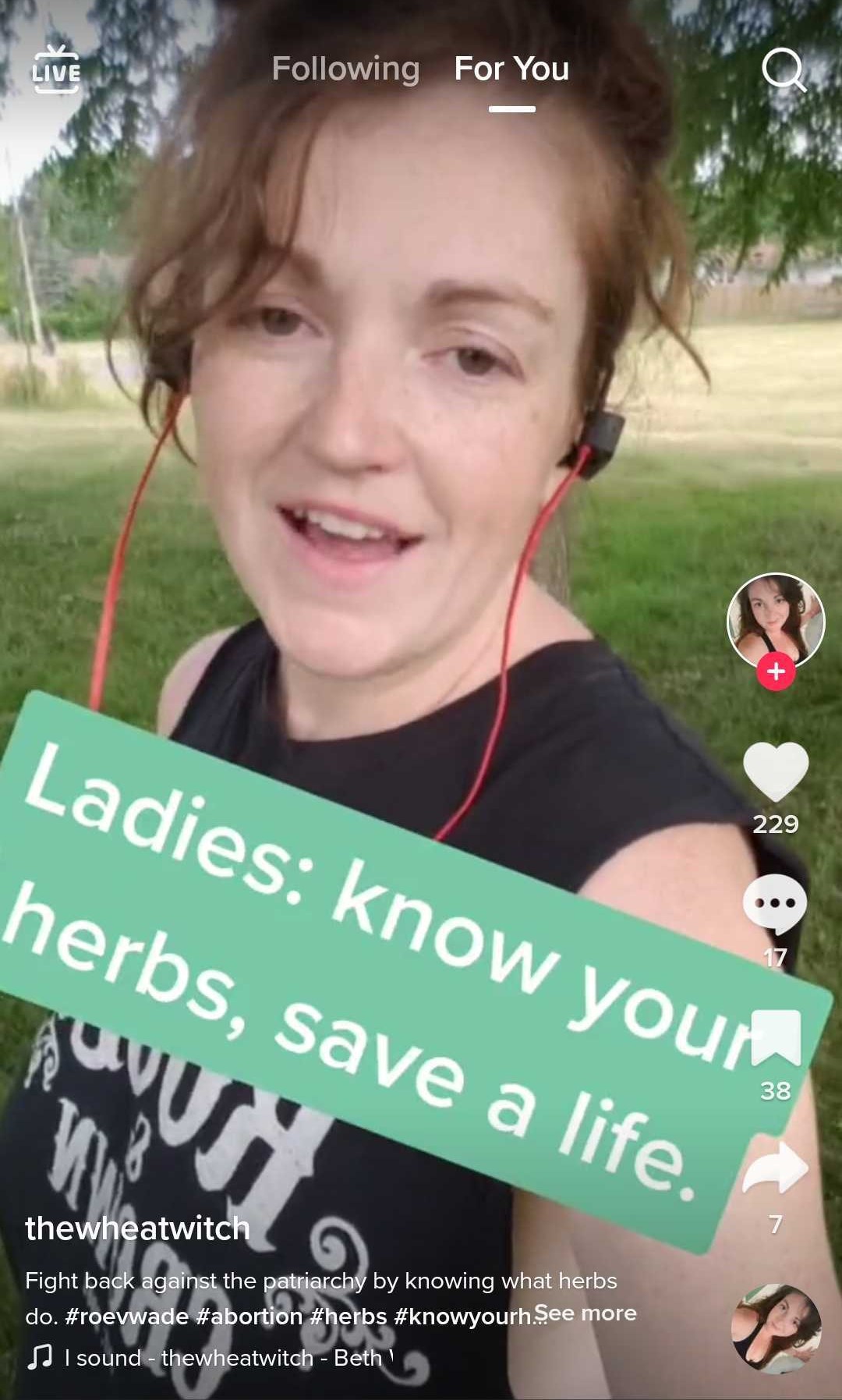}
  \caption{TikTok Post \#1}
  \label{fig:Post1}
\end{figure}

The first post we presented to participants was labeled by the creator with the hashtags \#roevwade, \#abortion, \#herb,s \#knowyourherbs, \#herbalist, \#womensrights, \#fightbackwithherbs, \#herbalism, \#homesteadinglife, \#michigan, \#crazyplantlady, and \#farmlife. This post discusses the use of \textit{Eastern Daisy Fleabane} root \citep{Thoms}. The use of fleabane as an abortifacient herbal tea was found misleading as it can ``have unpredictable effects'' and there is no evidence that this root can induce a miscarriage, as shown in Table \ref{tab:herbs}. The screenshot of the post as it appeared in the standard TikTok app is shown in Figure \ref{fig:Post1}. 

\subsubsection{Assessment}
We broke down the results of the participants' evaluation in two groups based on their baseline mental model of misinformation on TikTok we outlined in section \ref{sec:conceptualzation} above. The assessment results of the first TikTok video in our study are given in Table \ref{tab:post1-assessment}. Out of the 30 participants who thought misinformation is disseminated on TikTok without intent, 14 were randomly selected to assess the first post with abortion misinformation. Three of them thought the video is indeed misinformation, stating that ``\textit{this is likely misinformation, as a claim such as this likely has very little evidence to substantiate it}'' [\textbf{P46MM30}]. A surprising 50\% of the participants in this group though the video was not misinformation, feeling that ``\textit{it is true because she sound [sic] like she knows what she is talking about}'' [\textbf{P8FL30}]. Four participants in this group were unsure if this video was misinformation, worried that ``\textit{they state some historical context not verified in any way}'' [\textbf{P42FL30}].     

\begin{table}[htbp]
\renewcommand{\arraystretch}{1.3}
\footnotesize
\caption{Is Post \#1 Misinformation?}
\label{tab:post1-assessment}
\centering
\aboverulesep=0ex 
   \belowrulesep=0ex 
\begin{tabularx}{0.8\linewidth}{|Y|}
\Xhline{3\arrayrulewidth}
\toprule
 \textbf{Misinformation (no intent)} [\textbf{viewed}: 14 participants] \\\Xhline{3\arrayrulewidth}
\midrule
\footnotesize
\vspace{0.2em}
    \hfill \makecell{\textbf{Yes} \\ 3 (21.43\%)} 
    \hfill \makecell{\textbf{No} \\ 7 (50\%)} 
    \hfill \makecell{\textbf{Unsure} \\ 4 (28.57\%)} \hfill\null
\vspace{0.2em}
\\\Xhline{3\arrayrulewidth}
\midrule
 \textbf{Disinformation (intent)} [\textbf{viewed}: 12 participants] \\\Xhline{3\arrayrulewidth}
\midrule
\footnotesize
\vspace{0.2em}
    \hfill \makecell{\textbf{Yes} \\ 4 (33.34\%)} 
    \hfill \makecell{\textbf{No} \\ 3 (25\%)} 
    \hfill \makecell{\textbf{Unsure} \\ 5 (41.67\%)} \hfill\null
\vspace{0.2em}
\\\Xhline{3\arrayrulewidth}
\bottomrule
\end{tabularx}
\end{table}

Out of the other 30 participants that saw misinformation being spread with intent on TikTok, 12 were randomly shown the first post (the random selection was done by the Qualtrics survey software we used, leading to slightly unbalanced group). Four of them confirmed the video contains falsehoods with quite a verbose justification: ``\textit{This is misinformation. It is referencing clearing your liver which immediately points to something more like a Multi-Level Marketing (MLM) product; She's using the same terminology that essential oil salespeople use; These people can never name what toxins, etc you are eliminating because that is not actually happening}'' [\textbf{P50FL60}]. Three participants thought otherwise, believing this post was not misinformation because ``\textit{the content creator seems informed}'' [\textbf{P10MA40}]. Five participants were unsure, because they had ``\textit{no idea whether the claim in the video is based in truth or not}'' [\textbf{P32NC50}]. 

\subsubsection{Response} 
Participants were also asked to describe what action they would take for each post, with their actions given in Table \ref{tab:post1-action}. Of the participants that thought misinformation is disseminated on TikTok without intent, six (42.86\%) said that they would ``\textit{scroll past without interacting with the post}'' [\textbf{P42FL30}]. Participants in this group were equally likely to ``\textit{verify said information to see if it is accurate}'' [\textbf{P20FL40}] and ``\textit{would like the post}'' [\textbf{P8FL20}]. Only two (14.28\%) participants in this group said they would ``\textit{unfollow this person}'' [\textbf{P56FM50}] or ``\textit{report this video for dangerous activities}'' [\textbf{P24FR20}]. Of the participants who saw misinformation being spread with intent on TikTok, five (41.67\%) said they ``\textit{would just scroll past it}'' [\textbf{P10MA40}] and five (41.67\%) said they would ``\textit{look at the comments and then conduct my own personal research}'' [\textbf{P58FM30}]. There were also two (16.67\%) participants that said they ``\textit{would block it}'' [\textbf{P4MM50}] or ``\textit{would report this}''[\textbf{P50FL60}] and no participants from this group said they would like the post.

\begin{table}[htbp]
\renewcommand{\arraystretch}{1.3}
\footnotesize
\caption{What action would you take on Post \#1?}
\label{tab:post1-action}
\centering
\aboverulesep=0ex 
   \belowrulesep=0ex 
\begin{tabularx}{0.8\linewidth}{|Y|}
\Xhline{3\arrayrulewidth}
\toprule
 \textbf{Misinformation (no intent)} [\textbf{viewed}: 14 participants] \\\Xhline{3\arrayrulewidth}
\midrule
\footnotesize
\vspace{0.2em}
    \hfill \makecell{\textbf{Ignore} \\ 6 (42.86\%)} 
    \hfill \makecell{\textbf{Fact-check} \\ 3 (21.43\%)} 
    \hfill \makecell{\textbf{Block} \\ 1 (7.14\%)} 
    \hfill \makecell{\textbf{Report} \\ 1 (7.14\%)}
    \hfill \makecell{\textbf{Like} \\ 3 (21.43\%)} \hfill\null
\vspace{0.2em}
\\\Xhline{3\arrayrulewidth}
\midrule
 \textbf{Disinformation (intent)} [\textbf{viewed}: 12 participants] \\\Xhline{3\arrayrulewidth}
\midrule
\footnotesize
\vspace{0.2em}
    \hfill \makecell{\textbf{Ignore} \\ 5 (41.67\%)} 
    \hfill \makecell{\textbf{Fact-check} \\ 5 (41.67\%)} 
    \hfill \makecell{\textbf{Block} \\ 1 (8.33\%)} 
    \hfill \makecell{\textbf{Report} \\ 1 (8.33\%)}
    \hfill \makecell{\textbf{Like} \\ 0 (0\%)} \hfill\null
\vspace{0.2em}
\\\Xhline{3\arrayrulewidth}
\bottomrule
\end{tabularx}
\end{table}

\subsection{Post \#2}

\begin{figure}[htbp]
  \centering
  \includegraphics[width=0.35\linewidth]{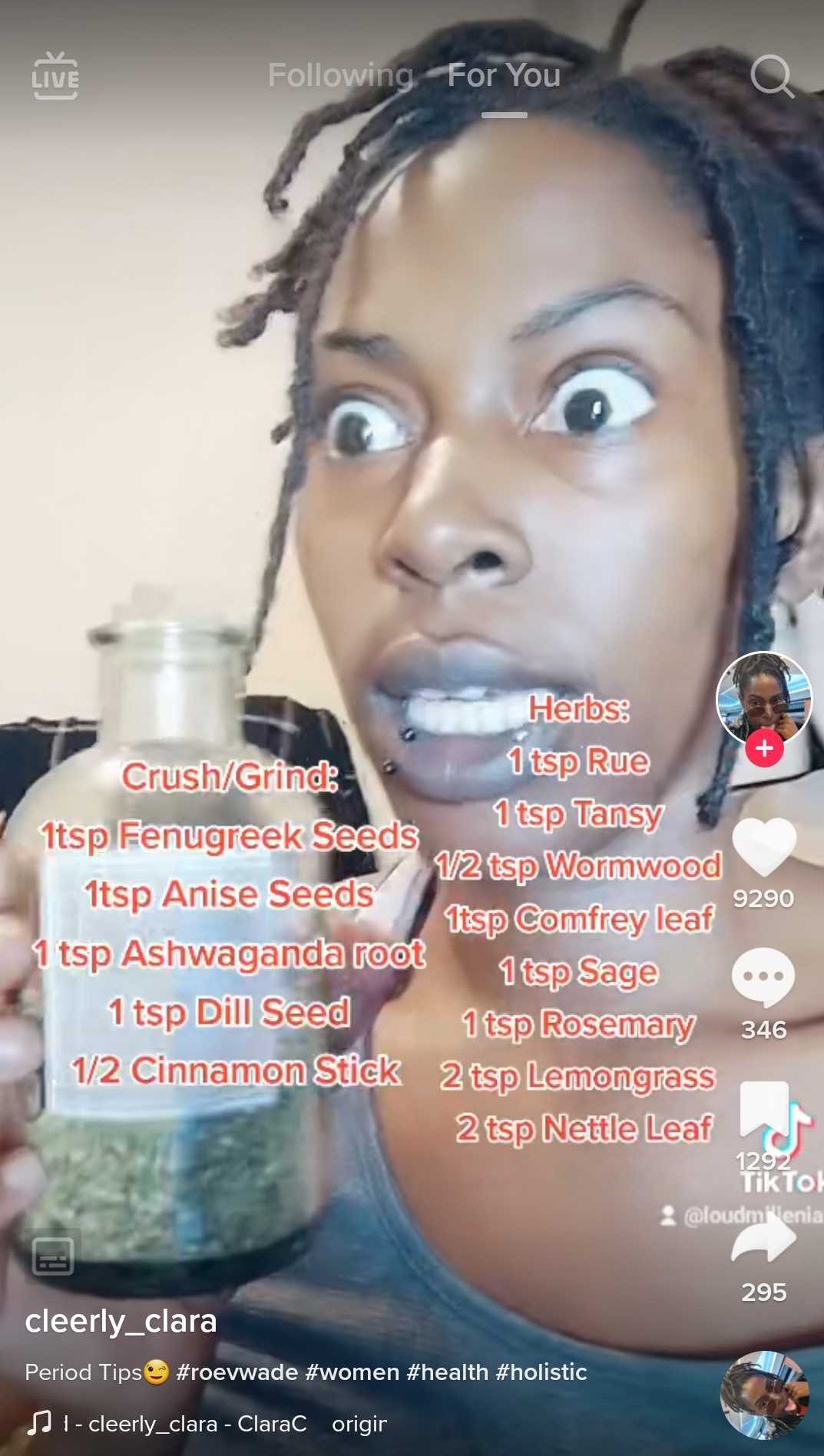}
  \caption{TikTok Post \#2}
  \label{fig:Post2}
\end{figure}

The second post that we presented to the participants was labeled by the creator with the hashtags \#roevwade, \#women, \#health, and \#holistic. This post discusses the use of an abortion tea containing multiple herbs, including rue, which has dangerous side effects noted in Table \ref{tab:herbs} and health advisories warn that it ``can lead to death for both the mother and baby'' \citep{Mosquera}. The screenshot of the post as it appeared in the standard TikTok app is shown in Figure \ref{fig:Post2}. 

\subsubsection{Assessment}
The participants that thought that misinformation is disseminated on TikTok without intent were almost evenly split regarding this post, as shown in Table \ref{tab:post2-assessment}. Six (46.15\%) said they ``\textit{don't believe this is misinformation}'' [\textbf{P33FL30}] and five (38.46\%) said they ``\textit{do think this post is misinformation}'' [\textbf{P13FR30}]. The remaining two (15.38\%) participants said they ``\textit{cannot confirm if this post is misinformation}'' [\textbf{P17FR30}]. Participants that saw falsehoods as disinformation on TikTok did not see this post as inaccurate as only two (15.38\%) of them thought the post is ``\textit{misinformation because it is suggesting that an herb blend is a safe and effective way to self administer an abortion}'' [\textbf{P36FL40}]. Five participants (38.46\%) said they ``\textit{don't think this is misinformation because the post provides full context on the information it was trying to provide}'' [\textbf{P40FL30}]. Six participants, or (46.15\%) were unsure because they ``\textit{don't have enough knowledge to know if it's misinformation}'' [\textbf{P59FL30}]. 

\begin{table}[htbp]
\renewcommand{\arraystretch}{1.3}
\footnotesize
\caption{Is Post \#2 Misinformation?}
\label{tab:post2-assessment}
\centering
\aboverulesep=0ex 
   \belowrulesep=0ex 
\begin{tabularx}{0.8\linewidth}{|Y|}
\Xhline{3\arrayrulewidth}
\toprule
 \textbf{Misinformation (no intent)} [\textbf{viewed}: 13 participants] \\\Xhline{3\arrayrulewidth}
\midrule
\footnotesize
\vspace{0.2em}
    \hfill \makecell{\textbf{Yes} \\ 5 (38.46\%)} 
    \hfill \makecell{\textbf{No} \\ 6 (46.15\%)} 
    \hfill \makecell{\textbf{Unsure} \\ 2 (15.38\%)} \hfill\null
\vspace{0.2em}
\\\Xhline{3\arrayrulewidth}
\midrule
 \textbf{Disinformation (intent)} [\textbf{viewed}: 13 participants] \\\Xhline{3\arrayrulewidth}
\midrule
\footnotesize
\vspace{0.2em}
    \hfill \makecell{\textbf{Yes} \\ 2 (15.38\%)} 
    \hfill \makecell{\textbf{No} \\ 5 (38.46\%)} 
    \hfill \makecell{\textbf{Unsure} \\ 6 (46.15\%)} \hfill\null
\vspace{0.2em}
\\\Xhline{3\arrayrulewidth}
\bottomrule
\end{tabularx}
\end{table}

\subsubsection{Response}
The 13 participants who thought of no intent behind misinformation on TikTok indicated they would perform a wide variety of activities for this post as shown in Table \ref{tab:post2-action}. Four (30.77\%) of them said they would ``\textit{potentially do some research into the other herbs that they are not familiar with}'' [\textbf{P17FR30}], three (23.08\%) said they ``\textit{would move past it}`` [\textbf{P8FL20}], two (15.38\%) said they ``\textit{would most likely block this account}'' [\textbf{P13FR30}], and two (15.38\%) said they ``\textit{would probably like this post}'' [\textbf{P31FL30}]. The two participants that said they would block this video indicated they felt very strongly about the content as it is ``\textit{definitely misinformation because there is scientific evidence that home remedies for this almost never work}'' [\textbf{P25FL20}] and ``\textit{it was extremely uncalled for in suggesting abortion tea; If this was in my TikTok feed I would [also] report this video}'' [\textbf{P24FR20}].

None of the participants that saw misinformation being spread with intent on TikTok said they would block or report this post. The most popular responses included that six (46.15\%) said they would ``\textit{simply move past the video and not pay no mind to it}'' [\textbf{P35FL30}] and five (38.46\%) said they would fact-check the content. \textbf{P5FL30} included that to determine the post is not misinformation they would need to see ``\textit{a Gynocologist [sic] [to] validate or debunk this information before I could trust it. This would include discussing the ingredients, linking relevant papers and studies, and reminding people to go see their doctor for personal medical information.}'' Two (15.38\%) of the participants indicated responses that they've ``\textit{done research on natural remedies for several health issues and if it was on my feed I may comment or like it}'' [\textbf{P6FA30}].

\begin{table}[htbp]
\renewcommand{\arraystretch}{1.3}
\footnotesize
\caption{What action would you take on Post \#2?}
\label{tab:post2-action}
\centering
\aboverulesep=0ex 
   \belowrulesep=0ex 
\begin{tabularx}{0.8\linewidth}{|Y|}
\Xhline{3\arrayrulewidth}
\toprule
 \textbf{Misinformation (no intent)} [\textbf{viewed}: 13 participants] \\\Xhline{3\arrayrulewidth}
\midrule
\footnotesize
\vspace{0.2em}
    \hfill \makecell{\textbf{Ignore} \\ 3 (23.08\%)} 
    \hfill \makecell{\textbf{Fact-check} \\ 4 (30.77\%)} 
    \hfill \makecell{\textbf{Block} \\ 2 (15.38\%)}
    \hfill \makecell{\textbf{Report} \\ 2 (15.38\%)}
    \hfill \makecell{\textbf{Like} \\ 2 (15.38\%)} \hfill\null
\vspace{0.2em}
\\\Xhline{3\arrayrulewidth}
\midrule
 \textbf{Disinformation (intent)} [\textbf{viewed}: 13 participants] \\\Xhline{3\arrayrulewidth}
\midrule
\footnotesize
\vspace{0.2em}
    \hfill \makecell{\textbf{Ignore} \\ 6 (46.15\%)} 
    \hfill \makecell{\textbf{Fact-check} \\ 5 (38.46\%)} 
    \hfill \makecell{\textbf{Block} \\ 0 (0\%)} 
    \hfill \makecell{\textbf{Report} \\ 0 (0\%)} 
    \hfill \makecell{\textbf{Like} \\ 2 (15.38\%)} \hfill\null
\vspace{0.2em}
\\\Xhline{3\arrayrulewidth}
\bottomrule
\end{tabularx}
\end{table}

\subsection{Post \#3}

\begin{figure}[htbp]
  \centering
  \includegraphics[width=0.35\linewidth]{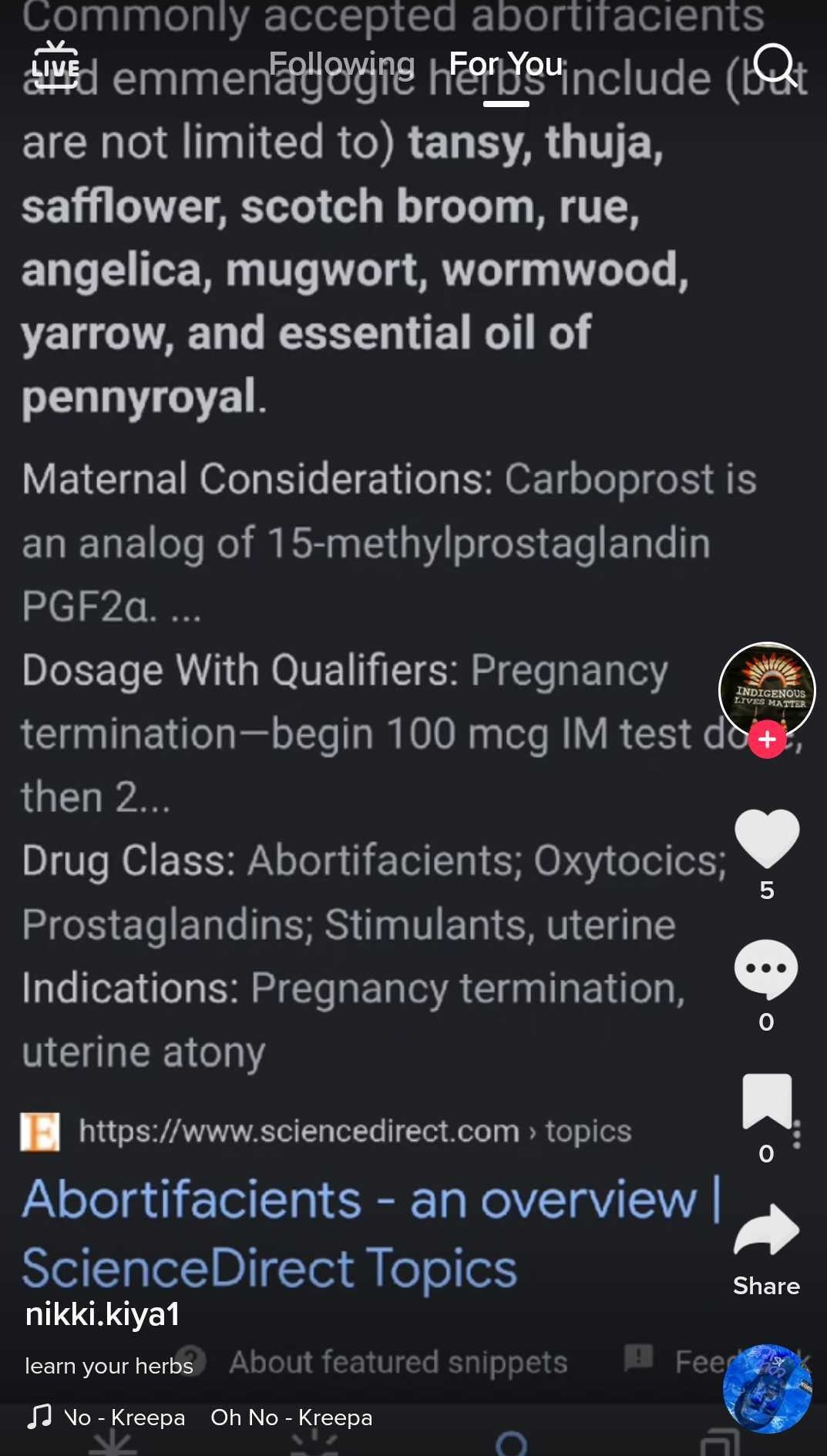}
  \caption{TikTok Post \#3}
  \label{fig:Post3}
\end{figure}

The third post that we presented to the participants appears to be a \textit{ScienceDirect} article indicating which herbs are abortifacients, but is actually a Google search result snippet with a caption stating ``\textit{learn your herbs}''. This TikTok video was not labeled with any hashtags by the creator. Although the full \textit{ScienceDirect} article purportedly referenced in this video provides warnings about the toxicity risks of the herbs \citep{Romm}, the screenprint also has directions on the dosage for pregnancy termination centrally positioned in the overall text. The screenshot, shown in Figure \ref{fig:Post3}, includes the pennyroyal and mugwort herbs  as abortifacients. 

\subsubsection{Assessment} Multiple participants in both groups observed that the TikTok post appeared to be from ScienceDirect. A few noticed it was a Google search result or that the ``\textit{source looks questionable}'' [\textbf{P49Fl30}] and reflected that in their responses. In the misinformation group, as shown in Table \ref{tab:post3-assessment}, six (40\%) of the participants said they ``\textit{think this post maybe fake because the whole information is not given just the negative information that they want you to see}'' [\textbf{P11ML40}]. Five (33.33\%) leaned towards it not being misinformation and  weighing that ``\textit{this post is presenting information from ScienceDirect, which I consider to be a reputable source of science-based and factual information}'' [\textbf{P44FL30}]. Four (26.67\%) participants were unsure because ``\textit{This post may contain some degree of accuracy given that the source is somewhat credible, but herbs are often not studied enough for these claims to be made with a high degree of accuracy}'' [\textbf{P46MM30}].

Most of the participants in the disinformation group were unsure if this post was misinformation. 10 (62.5\%) of them said they ``\textit{honestly cannot tell if this is misinformation; I see that the website credited is science based, but without going to the website, I am really unsure}'' [\textbf{P41FM40}]. The next most common response was that ``\textit{this tries to present as being medically accurate, but it's just a screenshot of Google search results, so I would not trust it on its face}'' [\textbf{P12FL60}]. Only \textbf{P15ML30} said he did not ``\textit{know if it's misinformation; I'd assume it isn't because I'm not how someone could lie about a google search coming up}.'' 

\begin{table}[htbp]
\renewcommand{\arraystretch}{1.3}
\footnotesize
\caption{Is Post \#3 Misinformation?}
\label{tab:post3-assessment}
\centering
\aboverulesep=0ex 
   \belowrulesep=0ex 
\begin{tabularx}{0.8\linewidth}{|Y|}
\Xhline{3\arrayrulewidth}
\toprule
 \textbf{Misinformation (no intent)} [\textbf{viewed}: 15 participants] \\\Xhline{3\arrayrulewidth}
\midrule
\footnotesize
\vspace{0.2em}
    \hfill \makecell{\textbf{Yes} \\ 6 (40\%)} 
    \hfill \makecell{\textbf{No} \\ 5 (33.33\%)} 
    \hfill \makecell{\textbf{Unsure} \\ 4 (26.67\%)} \hfill\null
\vspace{0.2em}
\\\Xhline{3\arrayrulewidth}
\midrule
 \textbf{Disinformation (intent)} [\textbf{viewed}: 16 participants] \\\Xhline{3\arrayrulewidth}
\midrule
\footnotesize
\vspace{0.2em}
    \hfill \makecell{\textbf{Yes} \\ 5 (31.25\%)} 
    \hfill \makecell{\textbf{No} \\ 1 (6.25\%)} 
    \hfill \makecell{\textbf{Unsure} \\ 10 (62.5\%)} \hfill\null
\vspace{0.2em}
\\\Xhline{3\arrayrulewidth}
\bottomrule
\end{tabularx}
\end{table}

\subsubsection{Response}
The participants in both groups, as indicated in Table \ref{tab:post3-action}, were mostly inclined to ignore this post, saying they ``\textit{wouldn't interact with such a post}'' [\textbf{P3ML50}] with \textbf{P29FL60} noting that the post ``\textit{was a standard google search so it may be true but again I would not respond and I would swipe on by}.'' The next most common response of the participants in the misinformation group was to ``\textit{report this creator}'' [\textbf{P16FL30}]. The remaining two (13.33\%) participants in this group said they would ``\textit{likely look at the comments to see what others have said}'' [\textbf{P19FL30}]. Six (37. 5\%) of the participants in the disinformation group said they ``\textit{would probably do research on the specifics if it was something I desired to learn more about}'' [\textbf{P40FL30}].  Only one participant in this group said they would block the post ``\textit{because it does not state the risks of using these}'' [\textbf{P34MM30}].

\begin{table}[htbp]
\renewcommand{\arraystretch}{1.3}
\footnotesize
\caption{What action would you take on Post \#3?}
\label{tab:post3-action}
\centering
\aboverulesep=0ex 
   \belowrulesep=0ex 
\begin{tabularx}{0.8\linewidth}{|Y|}
\Xhline{3\arrayrulewidth}
\toprule
 \textbf{Misinformation (no intent)} [\textbf{viewed}: 15 participants] \\\Xhline{3\arrayrulewidth}
\midrule
\footnotesize
\vspace{0.2em}
    \hfill \makecell{\textbf{Ignore} \\ 9 (60\%)} 
    \hfill \makecell{\textbf{Fact-check} \\ 2 (13.33\%)} 
    \hfill \makecell{\textbf{Block} \\ 0 (0\%)} 
    \hfill \makecell{\textbf{Report} \\ 4 (26.67\%)} 
    \hfill \makecell{\textbf{Like} \\ 0 (0\%)} \hfill\null
\vspace{0.2em}
\\\Xhline{3\arrayrulewidth}
\midrule
 \textbf{Disinformation (intent)} [\textbf{viewed}: 16 participants] \\\Xhline{3\arrayrulewidth}
\midrule
\footnotesize
\vspace{0.2em}
    \hfill \makecell{\textbf{Ignore} \\ 9 (56.25\%)} 
    \hfill \makecell{\textbf{Fact-check} \\ 6 (37.5\%)} 
    \hfill \makecell{\textbf{Block} \\ 1 (6.25\%)} 
    \hfill \makecell{\textbf{Report} \\ 0 (0\%)} 
    \hfill \makecell{\textbf{Like} \\ 0 (0\%)} \hfill\null
\vspace{0.2em}
\\\Xhline{3\arrayrulewidth}
\bottomrule
\end{tabularx}
\end{table}

\subsection{Post \#4}

\begin{figure}[htbp]
  \centering
  \includegraphics[width=0.35\linewidth]{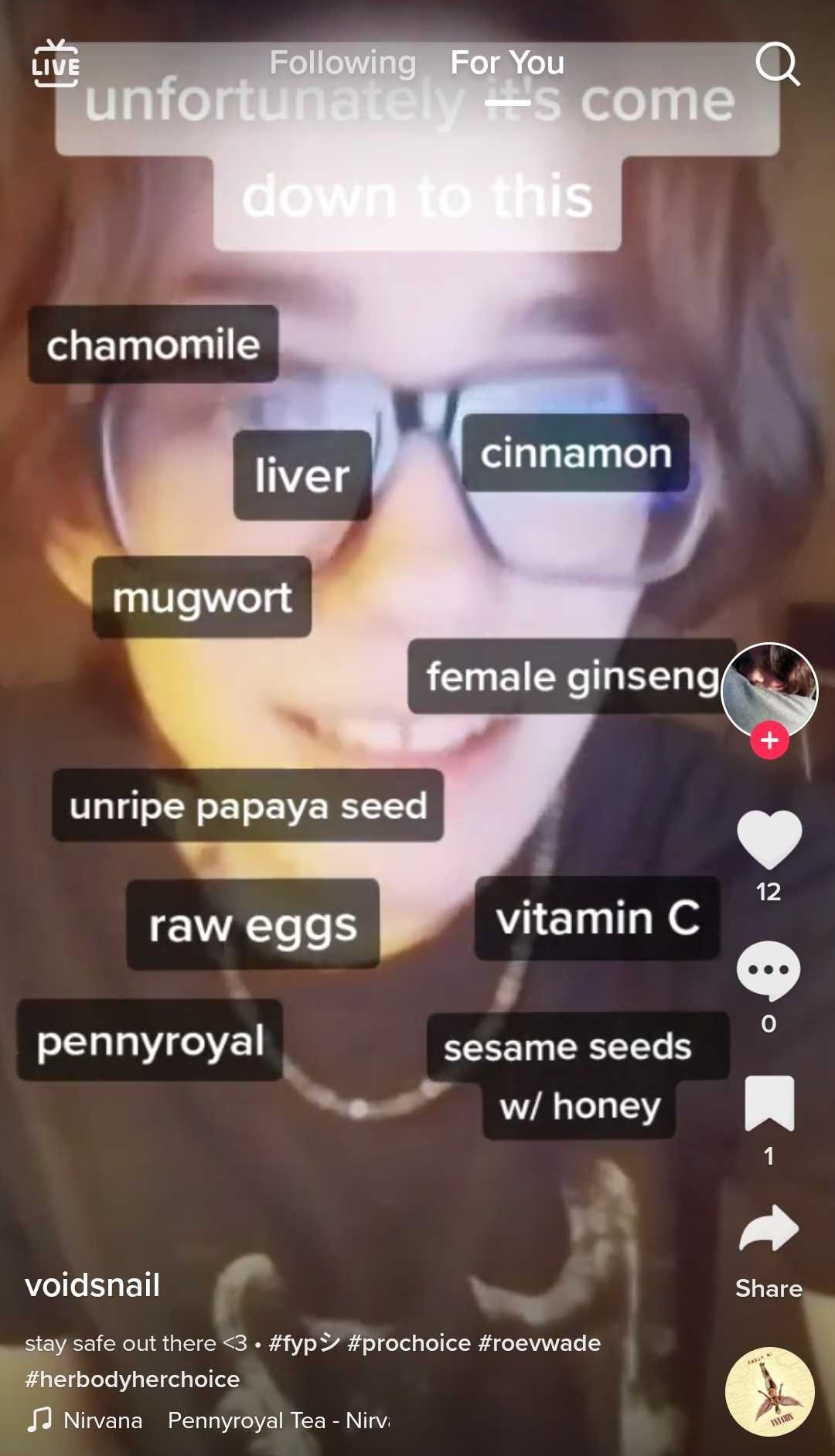}
  \caption{TikTok Post \#4}
  \label{fig:Post4}
\end{figure}

The fourth post presented to the participants was labeled by the creator with the hashtags \#fyp, \#prochoice, \#roevwade, and \#herbodyherchoice. The text overlay in the video also includes pennyroyal and mugwort under the heading ``\textit{unfortunately it's come down to this}''. The screenshot of the post as it appeared in the standard TikTok app is shown in Figure \ref{fig:Post4}. 

\subsubsection{Assessment}
As shown in Table \ref{tab:post4-assessment}, six (37.5\%) participants in the misinformation group indicated that they ``\textit{do think that this post is misinformation}'' [\textbf{P13FR30}], five (31.25\%) said it is ``\textit{not misinformation because it didn't advise a particular viewpoint or way of thinking}'' [\textbf{P54FL30}], and five (31.25\%) said ``\textit{It’s not clear that the post is misinformation because it’s just a list of herbs, supplements, and foods}'' [\textbf{P49Fl30}]. The disinformation group of participants felt more strongly that this post was misinformation, with nine (69.23\%) indicating ``\textit{this one is completely lacking context or supporting information, so, yes, I would qualify it as misinformation}'' [\textbf{P12FL60}]. Three (23.08\%) were ``\textit{not sure it’s misinformation but it may cause unsafe things to occur if not posted with caution}'' [\textbf{P35FL30}]. Only \textbf{P39FL30} said she ``\textit{doesn't really think it's misinformation}''.

\begin{table}[htbp]
\renewcommand{\arraystretch}{1.3}
\footnotesize
\caption{Is Post \#4 Misinformation?}
\label{tab:post4-assessment}
\centering
\aboverulesep=0ex 
   \belowrulesep=0ex 
\begin{tabularx}{0.8\linewidth}{|Y|}
\Xhline{3\arrayrulewidth}
\toprule
 \textbf{Misinformation (no intent)} [\textbf{viewed}: 16 participants] \\\Xhline{3\arrayrulewidth}
\midrule
\footnotesize
\vspace{0.2em}
    \hfill \makecell{\textbf{Yes} \\ 6 (37.5\%)} 
    \hfill \makecell{\textbf{No} \\ 5 (31.25\%)} 
    \hfill \makecell{\textbf{Unsure} \\ 5 (31.25\%)} \hfill\null
\vspace{0.2em}
\\\Xhline{3\arrayrulewidth}
\midrule
 \textbf{Disinformation (intent)} [\textbf{viewed}: 13 participants] \\\Xhline{3\arrayrulewidth}
\midrule
\footnotesize
\vspace{0.2em}
    \hfill \makecell{\textbf{Yes} \\ 9 (69.23\%)} 
    \hfill \makecell{\textbf{No} \\ 1 (7.69\%)} 
    \hfill \makecell{\textbf{Unsure} \\ 3 (23.08\%)} \hfill\null
\vspace{0.2em}
\\\Xhline{3\arrayrulewidth}
\bottomrule
\end{tabularx}
\end{table}

\subsubsection{Response}
When asked to describe the action they would take on this post, 8 (50\%) participants in the misinformation group said they ``\textit{would just ignore the video and move on}'' [\textbf{P52MM40}], as shown in Table \ref{tab:post4-action}. Three (18.75\%) participants said they ``\textit{would probably read through the comments and either search for similar videos on TikTok or Google it}'' [\textbf{P33FL30}], another three (18.75\%) said they would ``\textit{report this creator because their information is dangerous}'' [\textbf{P16FL30}], and the remaining two (12.5\%) said they would ``\textit{most likely block this user}'' [\textbf{P13FR30}]. Almost all of the participants in the disinformation group would ignore the fourth post. 11 (84.62\%) said ``\textit{would probably scroll past this without interacting because even if it is truth it isn't helpful or informative}'' [\textbf{P48FL30}]. The remaining two participants said they ``\textit{would try to fact check as best as I could}'' [\textbf{P6FA30}] and ``\textit{report this post}'' [\textbf{P12FL60}]. No participants in this group said they would block this post and no participants in either group said they would like the post.

\begin{table}[htbp]
\renewcommand{\arraystretch}{1.3}
\footnotesize
\caption{What action would you take on Post \#4?}
\label{tab:post4-action}
\centering
\aboverulesep=0ex 
   \belowrulesep=0ex 
\begin{tabularx}{0.8\linewidth}{|Y|}
\Xhline{3\arrayrulewidth}
\toprule
 \textbf{Misinformation (no intent)} [\textbf{viewed}: 16 participants] \\\Xhline{3\arrayrulewidth}
\midrule
\footnotesize
\vspace{0.2em}
    \hfill \makecell{\textbf{Ignore} \\ 8 (50\%)} 
    \hfill \makecell{\textbf{Fact-check} \\ 3 (18.75\%)} 
    \hfill \makecell{\textbf{Block} \\ 2 (12.5\%)} 
    \hfill \makecell{\textbf{Report} \\ 3 (18.75\%)} 
    \hfill \makecell{\textbf{Like} \\ 0 (0\%)} \hfill\null
\vspace{0.2em}
\\\Xhline{3\arrayrulewidth}
\midrule
 \textbf{Disinformation (intent)} [\textbf{viewed}: 13 participants] \\\Xhline{3\arrayrulewidth}
\midrule
\footnotesize
\vspace{0.2em}
    \hfill \makecell{\textbf{Ignore} \\ 11 (84.62\%)} 
    \hfill \makecell{\textbf{Fact-check} \\ 1 (7.69\%)} 
    \hfill \makecell{\textbf{Block} \\ 0 (0\%)} 
    \hfill \makecell{\textbf{Report} \\ 1 (7.69\%)} 
    \hfill \makecell{\textbf{Like} \\ 0 (0.00\%)} \hfill\null
\vspace{0.2em}
\\\Xhline{3\arrayrulewidth}
\bottomrule
\end{tabularx}
\end{table}

\subsection{Post \#5}

\begin{figure}[htbp]
  \centering
  \includegraphics[width=0.35\linewidth]{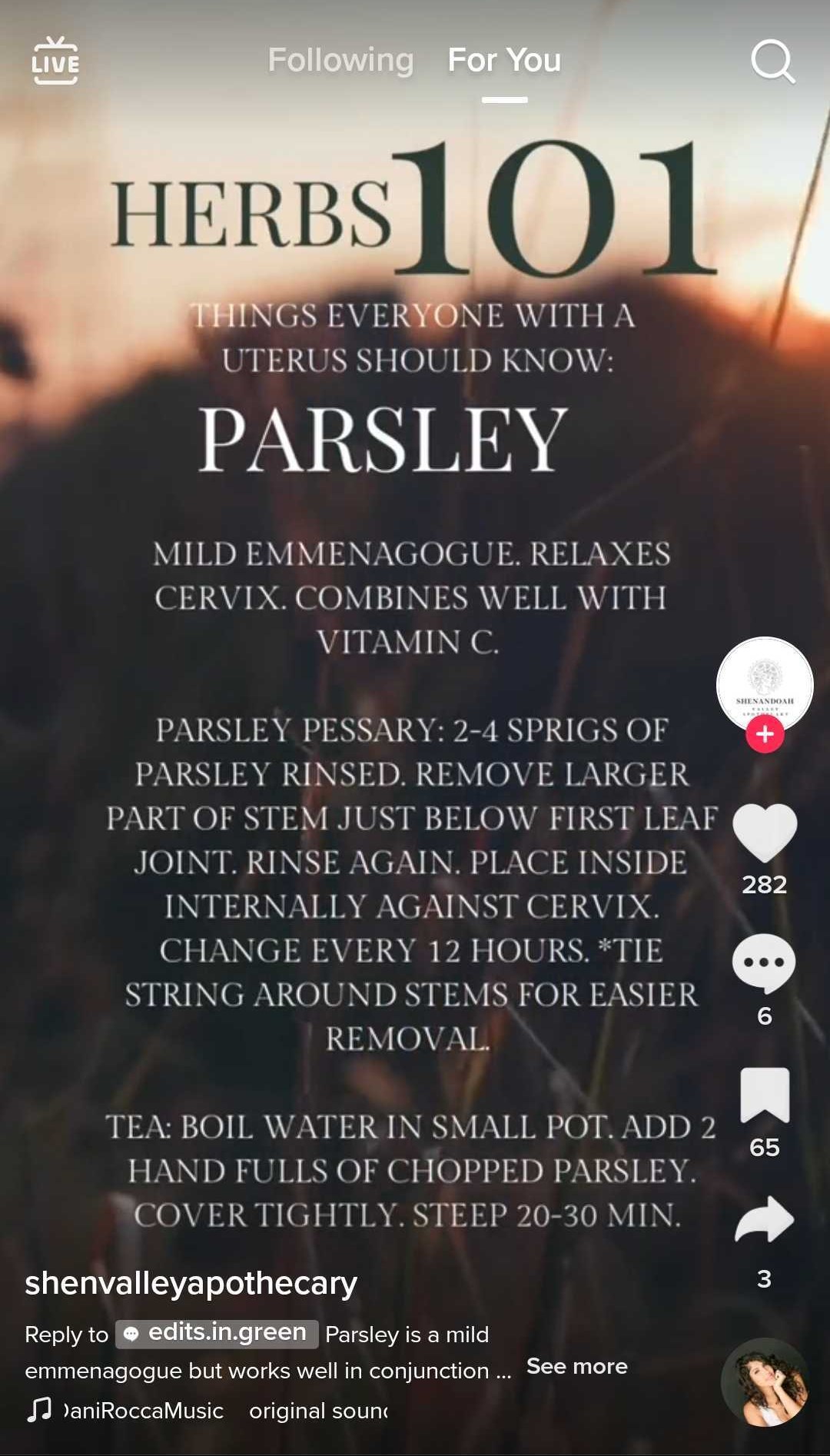}
  \caption{TikTok Post \#5}
  \label{fig:Post5}
\end{figure}

The fifth post that we presented was labeled with the hashtags \#themoreyouknow, \#parsley, \#pessaryinsertion, \#pessary, \#fertilityherbs, \#fertilityeducation, \#plantsheal, \#herbalist, \#apothecarycabinet, \#hippocraticoath, \#ayurvedic, and \#hippocratesfatherofmedicine. This creator has a series of posts that contain potential ``abortion inducers'' and emmenagogues (herbs which purpotedly stimulate menstruation). The post explains how to insert parsley into the cervix or make it into a tea. In 2018, a women in Argentina died from attempting to induce a miscarriage while utilizing this method, which ``stimulates blood flow in the uterus and can lead to massive internal bleeding and convulsions'' \citep{Cassella}. The screenshot of the post as it appeared in the standard TikTok app is shown in Figure \ref{fig:Post5}. 

\subsubsection{Assessment}
As shown in Table \ref{tab:post5-assessment}, most participants from both groups felt this post was misinformation. Twenty-two participants in total were randomly selected to assess this video. The groups were evenly distributed. Six (54.55\%) participants in the misinformation group stated they ``\textit{feel this post is misinformation because it does not relay the effects of placing an unsterile object inside your body, which is what the video is promoting/suggesting}'' [\textbf{P17FR30}] and that ``\textit{this post has no credible citations to support the claims it is making}'' [\textbf{P14FL30}]. Three (27.27\%) of the participants who thought misinformation is disseminated on TikTok without intent felt it was not misinformation because ``\textit{It’s always the consumers [sic] responsibility to do their own research and make their own choice(s)}'' [\textbf{P23FL30}]. Two (18.18\%) participants in this group said they ``\textit{have no idea}'' [\textbf{P21FA40}]. Eight (72.73\%) participants in the disinformation group thought ``\textit{it is misinformation because the account is not stating what the benefits of parsley even are, nor where they gathered this information}'' [\textbf{P7FM30}]. Three (27.27\%) said they ``\textit{suspect this is misinformation but do not know for sure}]'' [\textbf{P36FL40}]. No participants in the disinformation group thought the post wasn't misinformation.

\begin{table}[htbp]
\renewcommand{\arraystretch}{1.3}
\footnotesize
\caption{Is Post \#5 Misinformation?}
\label{tab:post5-assessment}
\centering
\aboverulesep=0ex 
   \belowrulesep=0ex 
\begin{tabularx}{0.8\linewidth}{|Y|}
\Xhline{3\arrayrulewidth}
\toprule
 \textbf{Misinformation (no intent)} [\textbf{viewed}: 11 participants] \\\Xhline{3\arrayrulewidth}
\midrule
\footnotesize
\vspace{0.2em}
    \hfill \makecell{\textbf{Yes} \\ 6 (54.55\%)} 
    \hfill \makecell{\textbf{No} \\ 3 (27.27\%)} 
    \hfill \makecell{\textbf{Unsure} \\ 2 (18.18\%)} \hfill\null
\vspace{0.2em}
\\\Xhline{3\arrayrulewidth}
\midrule
 \textbf{Disinformation (intent)} [\textbf{viewed}: 11 participants] \\\Xhline{3\arrayrulewidth}
\midrule
\footnotesize
\vspace{0.2em}
    \hfill \makecell{\textbf{Yes} \\ 8 (72.73\%)} 
    \hfill \makecell{\textbf{No} \\ 0 (0\%)} 
    \hfill \makecell{\textbf{Unsure} \\ 3 (27.27\%)} \hfill\null
\vspace{0.2em}
\\\Xhline{3\arrayrulewidth}
\bottomrule
\end{tabularx}
\end{table}

\subsubsection{Response}
The misinformation group participants, as shown in Table \ref{tab:post5-action} primarily said they ``\textit{would ignore this post}'' [\textbf{P14FL30}]. Two (18.18\%) participants said they would ``\textit{report the video as unsafe due to the major consequences that doing what the video says to do could ensue on someone's health}'' [\textbf{P17FR30}], one (9.09\%) said they ``\textit{would like to learn more and can always compare information on google}'' [\textbf{P11ML40}], and one (9.09\%) said they ``\textit{would most likely block this account}'' [\textbf{P13FR30}]. Participants in the disinformation group were mostly split on the post's inaccuracies, saying they ``\textit{wouldn't respond and just keep scrolling}'' [\textbf{P26FL30}] and they ``\textit{would look up the benefits of parsley to see if the post has an validation}'' [\textbf{P6FA30}]. Two (18.18\%) participants said they would ``\textit{typically report this post}`` [\textbf{P7FM30}] and \textbf{P38FL30} said she would ``\textit{block this account}.'' No participants in either group said they would like this video.

\begin{table}[htbp]
\renewcommand{\arraystretch}{1.3}
\footnotesize
\caption{What action would you take on Post \#5?}
\label{tab:post5-action}
\centering
\aboverulesep=0ex 
   \belowrulesep=0ex 
\begin{tabularx}{0.8\linewidth}{|Y|}
\Xhline{3\arrayrulewidth}
\toprule
 \textbf{Misinformation (no intent)} [\textbf{viewed}: 11 participants] \\\Xhline{3\arrayrulewidth}
\midrule
\footnotesize
\vspace{0.2em}
    \hfill \makecell{\textbf{Ignore} \\ 7 (63.74\%)} 
    \hfill \makecell{\textbf{Fact-check} \\ 1 (9.09\%)} 
    \hfill \makecell{\textbf{Block} \\ 1 (9.09\%)} 
    \hfill \makecell{\textbf{Report} \\ 2 (18.18\%)} 
    \hfill \makecell{\textbf{Like} \\ 0 (0\%)} \hfill\null
\vspace{0.2em}
\\\Xhline{3\arrayrulewidth}
\midrule
 \textbf{Disinformation (intent)} [\textbf{viewed}: 11 participants] \\\Xhline{3\arrayrulewidth}
\midrule
\footnotesize
\vspace{0.2em}
    \hfill \makecell{\textbf{Ignore} \\ 4 (36.36\%)} 
    \hfill \makecell{\textbf{Fact-check} \\ 4 (36.36\%)} 
    \hfill \makecell{\textbf{Block} \\ 1 (9.09\%)} 
    \hfill \makecell{\textbf{Report} \\ 2 (18.18\%)}
    \hfill \makecell{\textbf{Like} \\ 0 (0\%)} \hfill\null
\vspace{0.2em}
\\\Xhline{3\arrayrulewidth}
\bottomrule
\end{tabularx}
\end{table}

\subsection{Post \#6}
\begin{figure}[htbp]
  \centering
  \includegraphics[width=0.35\linewidth]{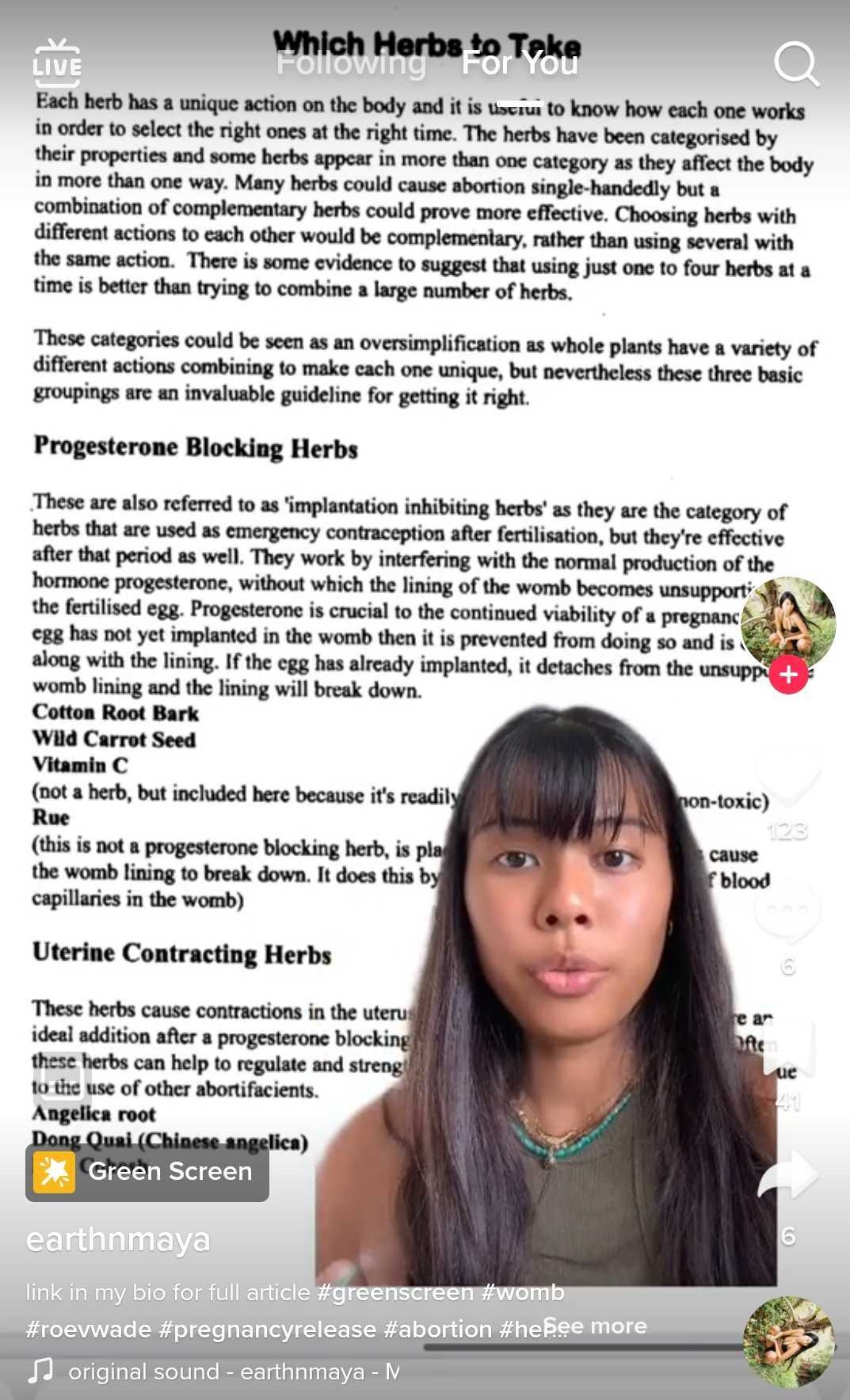}
  \caption{TikTok Post \#6}
  \label{fig:Post6}
\end{figure}

The sixth post that we presented participants was labeled with the hashtags \#greenscreen, \#womb, \#roevwade, \#pregnancyrelease, \#abortion, \#herbal, \#safety, and \#withlove. This post contains information on which herbs to use to perform a ``\textit{pregnancy release}''. The herbs in the video include rue, pennyroyal, and mugwort, which were also included in prior videos. The source cited in the video is indicated by the creator to be an article on herbal abortion from \textit{we.riseup.net}. The screenshot of the post as it appeared in the standard TikTok app is shown in Figure \ref{fig:Post6}. 

\subsubsection{Assessment}
As shown in Table \ref{tab:post6-assessment}, six (42.86\%) of the misinformation group participants indicated that this post was misinformation. Participant \textbf{P19FL30} explained that ``\textit{the content creator in this video does tell viewers to do `proper research' on the herbs she is promoting before using them; However, she notes that if `something goes wrong' and medical attention is needed that the viewer does not need to tell their medical provider that they have been taking these herbs - assuming that the creator is not a medical professional, I would consider this to be misinformation}.'' Five (35.71\%) of the participants said they weren't sure if this post is misinformation because ``\textit{there is an article supporting the statements made by the speaker and the speaker seems to have genuine intent; However the shared article is not in the form of a scientific verified and peer reviewed study}'' [\textbf{P42FL30}]. The remaining three participants in this group said they ``\textit{don't believe this post is misinformation because she provides supporting facts that you can double check and they will confirm her point}'' [\textbf{P54FL30}]. 

Six (35.29\%) of the participants in the disinformation group felt ``\textit{this is not misinformation because she is well spoken, shows where she is getting the information from, states what it is for and what it does and also mentions to seek medical help}'' [\textbf{P51FL40}]. Six (35.29\%) were unsure because they ``\textit{can't confirm if this is misinformation or not because I am not familiar with the scientific data that either backs up or refutes this info}''[\textbf{P1MR50}] and five (29.41\%) stated ``\textit{Yes, this post is misinformation; To start - the post screenshots a website: weriseup.net; That website is not a reliable medical source of information. I would scroll past}'' [\textbf{P41FM4}].

\begin{table}[htbp]
\renewcommand{\arraystretch}{1.3}
\footnotesize
\caption{Is Post \#6 Misinformation?}
\label{tab:post6-assessment}
\centering
\aboverulesep=0ex 
   \belowrulesep=0ex 
\begin{tabularx}{0.8\linewidth}{|Y|}
\Xhline{3\arrayrulewidth}
\toprule
 \textbf{Misinformation (no intent)} [\textbf{viewed}: 14 participants] \\\Xhline{3\arrayrulewidth}
\midrule
\footnotesize
\vspace{0.2em}
    \hfill \makecell{\textbf{Yes} \\ 6 (42.86\%)} 
    \hfill \makecell{\textbf{No} \\ 3 (21.43\%)} 
    \hfill \makecell{\textbf{Unsure} \\ 5 (35.71\%)} \hfill\null
\vspace{0.2em}
\\\Xhline{3\arrayrulewidth}
\midrule
 \textbf{Disinformation (intent)} [\textbf{viewed}: 17 participants] \\\Xhline{3\arrayrulewidth}
\midrule
\footnotesize
\vspace{0.2em}
    \hfill \makecell{\textbf{Yes} \\ 5 (29.41\%)} 
    \hfill \makecell{\textbf{No} \\ 6 (35.29\%)} 
    \hfill \makecell{\textbf{Unsure} \\ 6 (35.29\%)} \hfill\null
\vspace{0.2em}
\\\Xhline{3\arrayrulewidth}
\bottomrule
\end{tabularx}
\end{table}

\subsubsection{Response}
Table \ref{tab:post6-action} indicates that more that half of the participants from both groups said they ``\textit{would scroll past}'' [\textbf{P41FM40}], ``\textit{would simply ignore this post}'' [\textbf{P1MR50}]. Of the participants in the misinformation group, four (28.57\%) said they would ``\textit{encourage viewers to follow up with doing their own research and a notice that every woman's body still responds differently to methods}'' [\textbf{P54FL30}], two (14.29\%) said they ``\textit{would simply block the user and move on}'' [\textbf{P52MM40}], and one (7.14\%) said they ``\textit{would report this post}'' [\textbf{P25FL20}]. The participants in the disinformation group did not say they would block this post. Three (17.65\%) of them indicated they would ``\textit{read the full article this video links, and find other videos to make a better judgement on the trueness of this post}'' [\textbf{P5FL30}] and two (11.76\%) said ``\textit{there are natural remedies that are legitimate but telling people they can take an herb for contraception is dangerous; I would report this one}[\textbf{P50FL60}]. No participants in either group said they would like this post.

\begin{table}[htbp]
\renewcommand{\arraystretch}{1.3}
\footnotesize
\caption{What action would you take on Post \#6?}
\label{tab:post6-action}
\centering
\aboverulesep=0ex 
   \belowrulesep=0ex 
\begin{tabularx}{0.8\linewidth}{|Y|}
\Xhline{3\arrayrulewidth}
\toprule
 \textbf{Misinformation (no intent)} [\textbf{viewed}: 14 participants] \\\Xhline{3\arrayrulewidth}
\midrule
\footnotesize
\vspace{0.2em}
    \hfill \makecell{\textbf{Ignore} \\ 7 (50\%)} 
    \hfill \makecell{\textbf{Fact-check} \\ 4 (28.57\%)} 
    \hfill \makecell{\textbf{Block} \\ 2 (14.29\%)} 
    \hfill \makecell{\textbf{Report} \\ 1 (7.14\%)} 
    \hfill \makecell{\textbf{Like} \\ 0 (0\%)} \hfill\null
\vspace{0.2em}
\\\Xhline{3\arrayrulewidth}
\midrule
 \textbf{Disinformation (intent)} [\textbf{viewed}: 17 participants] \\\Xhline{3\arrayrulewidth}
\midrule
\footnotesize
\vspace{0.2em}
    \hfill \makecell{\textbf{Ignore} \\ 12 (70.59\%)} 
    \hfill \makecell{\textbf{Fact-check} \\ 3 (17.65\%)} 
    \hfill \makecell{\textbf{Block} \\ 0 (0\%)} 
    \hfill \makecell{\textbf{Report} \\ 2 (11.76\%)} 
    \hfill \makecell{\textbf{Like} \\ 0 (0\%)} \hfill\null
\vspace{0.2em}
\\\Xhline{3\arrayrulewidth}
\bottomrule
\end{tabularx}
\end{table}

\subsection{Post \#7}
\begin{figure}[htbp]
  \centering
  \includegraphics[width=0.35\linewidth]{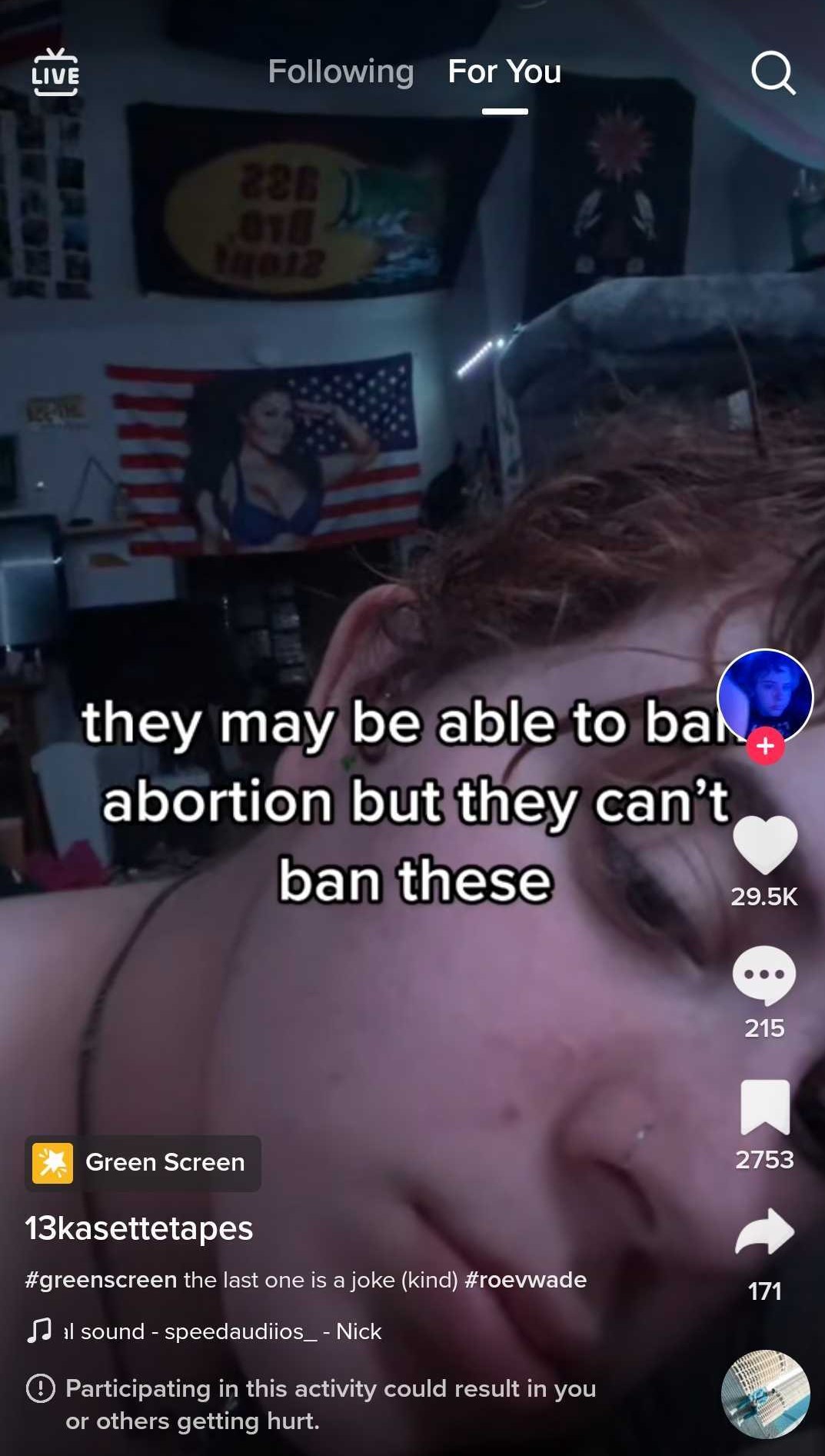}
  \caption{TikTok Post \#7}
  \label{fig:Post7}
\end{figure}

We also selected a post that was soft moderated by TikTok \citep{Keenan} and contained a trustworthiness tag with stating ``\textit{Participating in this activity could result in you or others getting hurt}.'' Post seven was labeled by the creator with the hashtages \#greenscreen and \#roevwade. As this is the only explicitly moderated post, we presented it to all 60 participants. The post, shown in Figure \ref{fig:Post7}, has overlay text that states ``\textit{they may be able to ban abortion but they can't ban these}'' and has slides of images that include pennyroyal, juniper berries, vitamin C, mugwort, aspirin, and a wire hanger.

\subsubsection{Assessment}
As shown in Table \ref{tab:post7-assessment}, 14 (46.67\%) of the 30 participants in the misinformation group said the post was misinformation ``\textit{considering this person is unqualified to be spreading this information}'' [\textbf{P16FL30}]. Eleven (36.67\%) said the post was not misinformation because they ``\textit{don't agree with using some of those products for achieving the goal that the video is promoting, but all of those products shown, used in the correct way/dosage can result in abortion}'' [\textbf{P17FR30}]. The remaining five (16.67\%) participants indicated they would ``\textit{have to research before deciding if it was misinformation or not}'' [\textbf{P44FL30}].

In the disinformation group, 13 (43.33\%) of participants indicated that ``\textit{this is misinformation as there is no evidence this is safe and effective means of self administering an abortion}'' [\textbf{P36FL40}]. Nine (30\%) of the participants said ``\textit{It is not helpful or informative; And the hanger is such a danger suggestion to put out there; I wouldn't consider this misinformation but it isn't helpful information}'' [\textbf{P48FL30}] and 8 (26.67\%) stated ``\textit{I don't have the scientific data to comment on whether or not this is misinformation}'' [\textbf{P1MR50}].

\begin{table}[htbp]
\renewcommand{\arraystretch}{1.3}
\footnotesize
\caption{Is Post \#7 Misinformation?}
\label{tab:post7-assessment}
\centering
\aboverulesep=0ex 
   \belowrulesep=0ex 
\begin{tabularx}{0.8\linewidth}{|Y|}
\Xhline{3\arrayrulewidth}
\toprule
 \textbf{Misinformation (no intent)} [\textbf{viewed}: 30 participants] \\\Xhline{3\arrayrulewidth}
\midrule
\footnotesize
\vspace{0.2em}
    \hfill \makecell{\textbf{Yes} \\ 14 (46.67\%)} 
    \hfill \makecell{\textbf{No} \\ 11 (36.67\%)} 
    \hfill \makecell{\textbf{Unsure} \\ 5 (16.67\%)} \hfill\null
\vspace{0.2em}
\\\Xhline{3\arrayrulewidth}
\midrule
 \textbf{Disinformation (intent)} [\textbf{viewed}: 30 participants] \\\Xhline{3\arrayrulewidth}
\midrule
\footnotesize
\vspace{0.2em}
    \hfill \makecell{\textbf{Yes} \\ 13 (43.33\%)} 
    \hfill \makecell{\textbf{No} \\ 9 (30\%)} 
    \hfill \makecell{\textbf{Unsure} \\ 8 (26.67\%)} \hfill\null
\vspace{0.2em}
\\\Xhline{3\arrayrulewidth}
\bottomrule
\end{tabularx}
\end{table}

\subsubsection{Response}
As indicated in Table \ref{tab:post7-action}, 13 (43.33\%) of the misinformation group and 18 (60\%) of the disinformation group said they ``\textit{wouldn't do anything with the post}'' [\textbf{P6FA30}] and they ``\textit{would not respond and just keep scrolling}'' [\textbf{P26FL30}]. Eleven participants of the misinformation group (36.67\%) and 8 of the disinformation group (26.67\%) said they ``\textit{would even consider reporting it as harmful because some people are viewing it and are not aware of how to properly use the products}'' [\textbf{P17FR30}] and ``\textit{because it's dangerous}'' [\textbf{P16FL30}]. Three (10\%) participants from each group said they would ``\textit{have to do more research}'' [\textbf{P2FL50}] and they ``\textit{would read the comments, look at similar videos, or Google it}'' [\textbf{P33FL30}]. Two (6.67\%) of the participants from the misinformation group said they would like the post as they ``\textit{agree with this post and have heard these methods before}'' [\textbf{P8FL20}]. One of participants from each group said they would block the post because ``\textit{it's not humorous joking around about women feeling they are being pushed to use alternative and possibly dangerous medicine}'' [\textbf{P60ML40}] and ``\textit{it is very disturbing and should not be allowed to be showed to young women in particular}'' [\textbf{P13FR30}].

\begin{table}[htbp]
\renewcommand{\arraystretch}{1.3}
\footnotesize
\caption{What action would you take on Post \#7?}
\label{tab:post7-action}
\centering
\aboverulesep=0ex 
   \belowrulesep=0ex 
\begin{tabularx}{0.8\linewidth}{|Y|}
\Xhline{3\arrayrulewidth}
\toprule
 \textbf{Misinformation (no intent)} [\textbf{viewed}: 30 participants] \\\Xhline{3\arrayrulewidth}
\midrule
\footnotesize
\vspace{0.2em}
    \hfill \makecell{\textbf{Ignore} \\ 13 (43.33\%)} 
    \hfill \makecell{\textbf{Fact-check} \\ 3 (10\%)} 
    \hfill \makecell{\textbf{Block} \\ 1 (3.33\%)} 
    \hfill \makecell{\textbf{Report} \\ 11 (36.67\%)} 
    \hfill \makecell{\textbf{Like} \\ 2 (6.67\%)} \hfill\null
\vspace{0.2em}
\\\Xhline{3\arrayrulewidth}
\midrule
 \textbf{Disinformation (intent)} [\textbf{viewed}: 30 participants] \\\Xhline{3\arrayrulewidth}
\midrule
\footnotesize
\vspace{0.2em}
    \hfill \makecell{\textbf{Ignore} \\ 18 (60\%)} 
    \hfill \makecell{\textbf{Fact-check} \\ 3 (10\%)} 
    \hfill \makecell{\textbf{Block} \\ 1 (3.33\%)} 
    \hfill \makecell{\textbf{Report} \\ 8 (26.67\%)} 
    \hfill \makecell{\textbf{Like} \\ 0 (0\%)} \hfill\null
\vspace{0.2em}
\\\Xhline{3\arrayrulewidth}
\bottomrule
\end{tabularx}
\end{table}

\section{Discussion} \label{sec:discussion}
Our first research question aimed to uncover how social media users conceptualize misinformation on TikTok, where does it originate, who are the targets, and what's its purpose. Misinformation, even within such a small sample as ours, invokes nuanced interpretations as people do not always stick to the popular ``fake news'' association \citep{Lewandowsky}. According to the participants in our sample, misinformation on TikTok might involve falsehoods only, but falsehoods could be interspersed with biased interpretations, facts taken out of context, or emotion-provoking narratives \citep{folk-models}. The amalgamation of factual and inaccurate content was mostly a craft assigned to individual content creators, often for personal gain, but the ``usual suspects'' -- political parties and foreign unfriendly countries to the United States -- were not spared in disseminating misinformation on Tiktok. 

This finding suggests that while TikTok's affordances are fit for selling products or participate in challenges \citep{Kaye}, they are not prohibitive for political and foreign actors to adapt their agenda setting and information operations. In the past, misleading narratives and content were migrating from fringe and alt-communities to the mainstream platforms \citep{ZannettouC} and the evidence from our study suggest that TikTok could be a future candidate destination. Our participants' view that many vulnerable people and easily-influenced crowds would unlikely step out from the ``For You'' page to check a claim adds to this impression, confirming previous studies that identified such users as the most sought out targets of misinformation \citep{Pennycook, Bronstein}. The TikTok participation model adds additional incentive to engage with misinformation as it provides immediate profit and long-term gains, according both to the participants in our sample and studies exploring political and conspirative influence on TikTok \citep{Medina-Serrano, Bleakley}.   

Our second research question narrowed our exploration on misleading abortion narratives. Given that we conducted our study after \textit{Roe vs Wade} was overturned, health misinformation regarding herbal abortifacients dominated the TikTok's `For You' pages of our participants. While some of them were simply intrigued about these ``at-home remedies,'' there were participants that were disturbed, worried, angered, and disgusted that dangerously misleading content like this finds its way on the platform. This emotion-provoking response, though concentrated on a topic of limited polarization, is worrisome because this is precisely the response that foreign actors sought to incite in past polarizing debates on social media \citep{Zannettou, Saeed}. These debates always involved politicization of the discourse, so the evidence that our participants also saw politically contextualized abortion narratives further suggest that TikTok might be the next health/political misinformation battleground \citep{Ghenai}, despite the impression of mostly apolitical participation on the platform so far \citep{Abidin}.  

But to substantiate such conjectures, one needs evidence on how actual users engage with essential, health-related abortion misinformation in the first place. Uncovering this evidence was central to our study and we attempted to gain as much as possible a nuanced insight into how users actually deal with various abortion misinformation content on TikTok. Our findings indicate that a  significant number of TikTok users do not see videos promoting herbal abortifacients as abortion misinformation, despite the scientific evidence to the contrary \citep{Gupta, Spencer, Godoy, Berglas, Patev, Rubin}. Regardless of whether the video included the creator itself, contained explicit tags to the herbal abortifacients listed in Table \ref{tab:herbs}, or was simply a textual post promoting these at-home remedies, many participants were unconvinced the posts were misinformation. The majority of them conceptualized misinformation as spreading falsehoods \textit{without} intent; those that saw intent behind the spread of falsehoods on TikTok, in general, were far more skeptical of the posts. 

A deeper look into the responses of the participants that did not see misinformation in the selected posts reveals that it was sufficient for the creators to appear ``like they know what they are talking about,'' ``provide context,'' ``seems informed,'' and are ``well spoken'' to make the misinformation believable. There was no particular demographic trend among these participants as the posts were equally convincing to all gender, political, and age groups. The group of participants that called out the misinformation indicated both analytical and heuristic cues that could be helpful for the wider audience on TikTok to employ in dealing with misleading abortion content. For example, videos that ``look like MLM promotions'', ``omit any side-effects'' of a proposed treatment, ``do not include credible citations'', and the ``creator is unqualified for making any health-related claims'' should be dismissed as abortion misinformation. All of these strategies have been proven to work against health misinformation in the past \citep{Pennycook-Rand-Psych}, so our findings strengthen the case for continuing the ``inoculation'' against false and misleading abortion narratives \citep{Lewandowsky}. 

This is especially important in situations where many users, like the ones in our study, remain unsure whether these misleading videos are abortion misinformation or not. Our results suggest that these users lack knowledge on the safety and the effectiveness of herbal abortifacients, which in turn precludes them to make a decisive dismissal of the misinformation. Educational interventions on TikTok regarding abortions have already be proposed \citep{Duggan}, but question is if the algorithmic curation of the `For You' page for TikTok users ``interested'' in alternatives includes them too. According to our findings, the undecided participants would be easily ``nudged'' with accurate scientific information, which is a strategy that worked for other types of misleading content \citep{Pennycook-accurate-nudges}. 


It is reassuring to see a fact-checking trend among even a small sample like ours. Perhaps affordances of TikTok probably encourage majority of the users to simply ignore many misleading posts, as many of the participants in our study did, and avoid to critically discern the content \citep{Pennycook1}. But recalling that familiarity, i.e. a repeated exposure to such videos makes the content look truthful \citep{Pennycook-Rand-Psych}, this fact-checking trend might need to be appended with better debunking and soft moderation efforts suggested to work for other health-related misinformation \citep{context2022, Swire-Thompson}. Such a need is further corroborated by our findings suggesting that even in the presence of the current TikTok soft moderation labels on post \#7, at least 30\% of the entire sample still believed that the content is \textit{not} misinformation. 

Equally important for consideration is that our results suggest a number of users are unafraid to block or report abortion misinformation. Prior evidence on flagging and reporting misleading content suggests that social media users might not opt for such actions as to preserve interpersonal relationships and avoid social arguments, especially for issue that do not matter much to them \citep{Tandoc}. However, reporting misinformation becomes a proactive endeavour when health-related misinformation is in question \citep{Bode}. Our results confirm that this effect also takes place for abortion-related misinformation and occurs across all demographics, making a further case for an actionable framework of misinformation sharing and correction sharing on TikTok \citep{Yu-Payton}.  

\subsection{Ethical Considerations}
The purpose of our study was not to generalize to a population; rather, to explore the phenomenon of individual dealing with abortion misinformation from a regular user perspective. While we added the participants' self-reported age, gender, and political leanings for more detailed reporting, we avoided providing definitive numbers accompanying the assessments and response strategies for each of the seven intervention posts with these demographics. Instead, we supported our findings with descriptive quotations from participants that convey the way ordinary users deal with misleading videos on TikTok in a hope that the results can help to elevate the study of abortion misinformation on social media as a whole. 

We acknowledge that there might be a potential risk of repeated exposure to abortion misinformation, i.e. an ``implied truth effect'' \citep{Pennycook-Rand-Psych}, as each participant saw four of the short-form videos. To mitigate this risk we explicitly pointed to the debunked information for the related ``at-home'' remedies and their associated harms. There is also a risk from oversimplification of our results where the participants' conceptualization, assessment, and response to abortion misinformation, expressed on their behalf, might not represent the entirety of the strategies used to deal with misleading videos. We, of course, agree that users do employ others ways and means to deal with misleading videos and we welcome every work that brings them to the fore. This will facilitate a scientific work on abortion misinformation beyond simple oppositional responses \citep{Southerton}, a corollary we also want to avoid when contextualizing our results for future anti-misinformation interventions.

\subsection{Limitations}
Our research was limited was limited in its scope to U.S. TikTok media users and the state of abortion misinformation regarding at-home remedies that existed on TikTok in the period immediately after the overturn of Roe vs Wade by the Supreme Court of the US. In as much as we attempted to have balanced, diverse, and inclusive set of short-form videos regarding herbal abortifacients, there certainly are, and will be, other similar content which our participants might assess and respond to differently than they did in our study. While the content of these videos was scientifically debunked and constituted misinformation during our study, we acknowledge that this might change in light of new scientific evidence. We are also limited by the state of the content moderation policies on TikTok that, together with public policy changes and new Supreme Court rulings, change and impact the relevance of our results, therefore we exercise caution in considering these results in the narrow post-Roe vs Wade context.

A limitation also comes from the sampling method and the use of a survey provider \citep{Redmiles2019}, as other users and other samples might provide results that differ from the once we obtained as there is little insight into general sampling and sample-related differences when users are broadly queried about abortion misinformation. By asking users directly about how they interact with abortion misinformation and misleading video content on TikTok, we got a wide variety of insights from a broad range of perspectives. We did not measure the efficacy of users' assessment approaches and response strategies for explicitly politicized abortion misinformation content, nor did we ask how users dealt with other abortion misinformation on other social media platforms. Short-form videos are a relatively new way of persuasive communication appealing to younger users, but traditional social media text, memes, and visceral images \citep{Basch} might provoke different responses for a wider population of users. Therefore, we are careful to avoid any predictive use of our findings because  TikTok's affordances might change in the future.

\section{Conclusion} \label{sec:conclusion}

Abortion misinformation undoubtedly shapes the way people make reproductive decisions and TikTok, a very popular social media platform, allows misleading content regarding ``at-home'' abortion remedies to reach wide audiences. Users possess the critical ability to assess, discern, and reject misleading and scientifically debunked abortion claims, but a worrying number of them are not ready to dismiss these alternatives for self-induced terminations of unwanted pregnancies in a post-\textit{Roe v Wade} America. Time will tell whether this proclivity for abortion misinformation will remain, but meanwhile we hope that TikTok, the scientific community, and health authorities take our results as actionable insight that can prevent harmful outcomes of abortion misinformation.

\bibliographystyle{ACM-Reference-Format}
\bibliography{sample-base}


\appendix

\section*{Study Questionnaire} \label{sec:survey}
\subsection*{Exposure and Preconceptions}
\begin{enumerate}
\itemsep 1em
        \item Can you please define ``misinformation'' in your own words (please be verbose): \textbf{[Open Ended]} 
        
        \item Have you encountered misinformation on TikTok and in what form? Please provide examples. \textbf{[Open Ended]}
        
        \item Where does misinformation on TikTok come from, in your opinion? \textbf{[Open Ended]}
        
        \item Who is the target of misinformation on TikTok, in your opinion? \textbf{[Open Ended]}
        
        \item Who benefits from misinformation on TikTok, in your opinion? \textbf{[Open Ended]}
\end{enumerate}

\subsection*{Engagement Strategies}
\begin{enumerate}
\itemsep 1em
        \item How do you suspect or know that a certain TikTok post is a misinformation? Please elaborate. \textbf{[Open Ended]}
        
        \item What is your strategy for dealing with misinformation posts on TikTok? Please elaborate. \textbf{[Open Ended]}
        
        \item Have you used any engagement features (e.g. share, comment, like, follow) for misinformation posts on TikTok? If so, in what circumstances? \textbf{[Open Ended]}
        
        \item Have you used any action features (e.g. block, mute, report, unfollow) for misinformation posts on TikTok? If so, in what circumstances? \textbf{[Open Ended]}
        
        \item Have you talked about a particular TikTok misinformation post outside social media? If so, in what circumstances? \textbf{[Open Ended]}
        
\end{enumerate}

\subsection*{Abortion Misinformation Exposure}
\begin{enumerate}
\itemsep 1em
    \item On what other occasions you have encountered misinformation regarding abortion on TikTok? Please elaborate. \textbf{[Open Ended]}
    
    \item What was your response to this particular abortion misinformation on TikTok? Please elaborate. \textbf{[Open Ended]}
        
    \item Where else does abortion misinformation exists outside of TikTok, in your opinion? Please elaborate and share any experiences. \textbf{[Open Ended]}
    
\end{enumerate}

\end{document}